\documentclass[pdflatex,sn-mathphys-num]{sn-jnl}% Math and Physical Sciences Numbered Reference Style 
\usepackage{graphicx}%
\usepackage{multirow}%
\usepackage{amsmath,amssymb,amsfonts}%
\usepackage{amsthm}%
\usepackage{mathrsfs}%
\usepackage[title]{appendix}%
\usepackage{xcolor}%
\usepackage{textcomp}%
\usepackage{manyfoot}%
\usepackage{booktabs}%
\usepackage{algorithm}%
\usepackage{algorithmicx}%
\usepackage{algpseudocode}%
\usepackage{listings}%
\theoremstyle{thmstyleone}%
\theoremstyle{thmstyletwo}%

\theoremstyle{thmstylethree}%

\usepackage{tabularx}
\usepackage{makecell}
\usepackage{graphicx}   % for \resizebox
\usepackage{booktabs}   % for \toprule, \midrule, \bottomrule

\newcommand{\MIR}[0]{\texttt{MACE4IRmol}}
\newcommand{\MACE}[0]{\texttt{MACE}}

\begin{document}

\title[Article Title]{MACE4IRmol: An uncertainty-aware foundation model for molecular infrared spectroscopy}
%MACE4IR: a foundation model for predicting infrared spectra of molecular systems
%MACE4IR: A general-purpose machine learning model for molecular infrared spectroscopy
%Towards a foundation model for molecular infrared spectroscopy

\author[1,2,3,4]{\fnm{Nitik} \sur{Bhatia}}\email{nitik.bhatia@tum.de}
\author[2,5,6]{\fnm{Ond\v{r}ej} \sur{Krej\v{c}\'{i}}}\email{ondrej.krejci@aalto.fi}

\author[7]{\fnm{Silvana} \sur{Botti}}\email{silvana.botti@rub.de}
\author*[1,2,3,4]{\fnm{Patrick} \sur{Rinke}}\email{patrick.rinke@tum.de}
\author*[7]{\fnm{Miguel} A.~L. \sur{Marques}}\email{miguel.marques@rub.de}

\affil*[1]{\orgdiv{Department of Physics}, \orgname{Technical University of Munich}, \orgaddress{\street{James-Franck-Strasse 1}, \city{Garching}, \postcode{85748}, \country{Germany}}}

\affil[2]{\orgdiv{Department of Applied Physics}, \orgname{Aalto University}, \orgaddress{\street{P.O. Box 11000}, \city{AALTO}, \postcode{FI-00076}, \country{Finland}}}

\affil[3]{\orgname{Munich Center for Machine Learning (MCML)}, \orgaddress{\city{Munich}, \country{Germany}}}

\affil[4]{\orgdiv{Atomistic Modelling Center}, \orgname{Munich Data Science Institute, Technical University of Munich}, \orgaddress{\street{Walther-Von-Dyck Str. 10}, \city{Garching}, \postcode{85748}, \country{Germany}}}

\affil[5]{\orgname{Department of Mechanical and Materials Engineering}, \orgaddress{\street{Vesilinnantie 5}, \city{Turku}, \country{Finland}}}

\affil[6]{\orgdiv{Department of Chemistry and Material Science}, \orgname{Aalto University}, \orgaddress{\street{P.O. Box 11000}, \city{AALTO}, \postcode{FI-00076}, \country{Finland}}}

\affil*[7]{\orgdiv{Research Center Future Energy Materials and Systems of the University Alliance Ruhr and Interdisciplinary Centre for Advanced Materials Simulation}, \orgname{Ruhr University}, \orgaddress{\street{Universitätsstraße 150}, \city{Bochum}, \postcode{D-44801}, \country{Germany}}}

%\affil*[6]{\orgdiv{Research Center Future Energy Materials and Systems of the University Alliance Ruhr and Interdisciplinary Centre for Advanced Materials Simulation}, \orgname{Ruhr University}, \orgaddress{\street{Universitätsstraße 150}, \city{Bochum}, \postcode{D-44801}, \country{Germany}}}

%\affil[3]{\orgdiv{Department}, \orgname{Organization}, \orgaddress{\street{Street}, \city{City}, \postcode{610101}, \state{State}, \country{Country}}}


\abstract{
Machine-learned interatomic potentials (MLIPs) have shown significant promise in predicting infrared spectra with high fidelity. However, the absence of general-purpose MLIPs that simultaneously span broad chemical diversity and provide reliable uncertainty estimates has limited their wider applicability. In this work, we introduce \MIR, an uncertainty-aware foundation model ensemble built on the \MACE{} architecture. \MIR{} is trained on \(\sim\)16 million molecular geometries and the corresponding density-functional theory (DFT) energies, forces, and dipole moments from the QCML dataset. The training data encompasses approximately 80 elements and a diverse set of molecules, including organic and inorganic compounds, and metal complexes. 
Importantly, \MIR{} is formulated as an ensemble of models to enable uncertainty quantification, which helps improve robustness in chemically diverse systems. Within this ensemble, separate models are trained with and without explicit dispersion corrections, allowing systematic assessment of van der Waals effects. In addition, \MIR{} delivers accurate predictions of energies, forces, dipole moments, and infrared spectra at a fraction of the computational cost of DFT, while enabling the explicit inclusion of nuclear quantum effects in infrared spectrum simulations. 
By combining generality, accuracy, efficiency, and uncertainty estimation, \MIR{} opens the door to rapid and reliable infrared spectra prediction for complex and diverse molecular systems.
}




\maketitle

\section{Introduction}\label{intro}

Infrared (IR) spectroscopy is a fundamental technique used to probe the vibrational properties of molecules and materials, offering valuable insight into molecular structures, bonding, and chemical dynamics \cite{kraack_ultrafast_2017}. IR spectroscopy plays a central role across diverse fields such as catalysis, drug design, environmental chemistry, and materials discovery \cite{Haas, Khan2018, C3CS60374A, molecules25122948, kraack_ultrafast_2017}. Accurately predicting IR spectra is crucial for interpreting experimental results, assigning spectral features, and accelerating the discovery and characterization of novel compounds \cite{Peter2015, lansford_infrared_2020, Interpret_exp_IR, Intepret_exp_IR_2}. 

Traditionally, IR spectra predictions have been based on first-principles approaches such as density functional theory (DFT), either within the harmonic approximation \cite{harmonic_1, Harmoni_2, AIMD_paper} or through \textit{ab initio} molecular dynamics (AIMD) \cite{AIMD_paper}. Harmonic analysis provides a computationally efficient means of estimating vibrational frequencies and intensities, but it overlooks anharmonicity and temperature-dependent effects, factors that are often essential for accurate spectral interpretation. In contrast, AIMD offers a more realistic representation by simulating vibrational dynamics at finite temperatures, thereby capturing these effects. Nonetheless, conventional AIMD simulations still treat nuclei classically and therefore neglect nuclear quantum effects (NQEs), such as zero-point energy and quantum delocalization, which can play a critical role in accurately reproducing IR spectra, particularly for light atoms and hydrogen-bonded systems \cite{Conte2023, ceriotti2010efficient}. These effects can be further incorporated into finite-temperature simulations through path integral molecular dynamics (PIMD) \cite{feynman1979path, feynman2018statistical}, which explicitly accounts for quantum nuclear statistics.

Across theoretical approaches, the computational cost of IR spectroscopy remains a central bottleneck. Addressing this challenge requires methods that reproduce near–first-principles accuracy in energies, forces, and dipole moments while remaining efficient enough to support the full range of applications, from harmonic analyses to long molecular dynamics simulations underlying classical or quantum dynamical treatments of nuclei.

Recent developments in machine-learned interatomic potentials (MLIPs) have significantly advanced the field by offering near first-principles accuracy at a fraction of the computational cost \cite{Behler_2007, bartok_gaussian_2010, smith_ani-1_2017, gastegger_machine_2017, SchNet, GM_2020, GM_2021, GPR, NequiP, Allegro}. Early MLIP frameworks, such as the Behler–Parrinello neural network potentials \cite{Behler_2007,behler_four_2021}, or kernel-based models like the Gaussian Approximation Potentials \cite{bartok_gaussian_2010}, demonstrated good accuracy and adaptability across diverse chemical and materials systems. More recently, the emergence of message-passing neural networks, particularly equivariant graph neural networks \cite{PaiNN, NequiP, Allegro, Batatia2022Design, MACE, GRACE}, has led to notable improvements in data efficiency, predictive accuracy, and model transferability. 
By embedding rotational and permutational symmetries, these architectures offer more physically grounded representations and generalize better to new chemical environments. Their shared representations and tensor operations also avoid combinatorial scaling with chemical species, thus allowing training highly multi-elemental models \cite{ACE, PaiNN, NequiP, Allegro, Batatia2022Design, MACE, GRACE}.

Despite rapid progress, most existing MLIPs are tailored to specific systems, such as organic molecules, nanoclusters, bulk crystals, or catalytic surfaces, and are typically trained on expensive DFT datasets curated for those specific applications \cite{PaiNN, lansford_infrared_2020, Flare_2, tang_machine_2023, ko_recent_2023, omranpour_machine_2025, olajide_application_2025, gurlek2025accuratemachinelearninginteratomic}. As a result, they often lack the flexibility to generalize beyond their original domain, limiting their transferability and reuse across different material or molecular systems \cite{montes_de_oca_zapiain_training_2022, ko_recent_2023, kandy_2023, olajide_application_2025}. In recent years, several initiatives have focused on developing more generalizable MLIPs by training them on large, chemically diverse datasets, with the goal of enhancing their performance across previously unseen chemical environments \cite{Chen_2022, deng2023chgnetpretraineduniversalneural, yang2024mattersimdeeplearningatomistic, yin2025alphanetscalinglocalframebasedatomistic, GRACE, kim2024dataefficientmultifidelitytraininghighfidelity, rhodes2025orbv3atomisticsimulationscale, fu2025learningsmoothexpressiveinteratomic, batatia2024foundationmodelatomisticmaterials, wood2025umafamilyuniversalmodels}. However, even the most general-purpose MLIPs are seldom developed with vibrational or spectroscopic accuracy in mind. Specifically for IR spectral prediction, which require not only precise energies and forces calculations but also accurate dipole moments, widely applicable MLIPs are lacking.

Recently, several MLIPs have incorporated dipole moment prediction capabilities, which enable both harmonic frequency analysis and molecular dynamics-based (AIMD and PIMD) spectral  \cite{grisafi_symmetry-adapted_2018, PhysNet, gastegger_machine_2021, PaiNN, beckmann_infrared_2022, MACE, tang_machine_2023, zou_deep_2023, stienstra_graphormer-ir_2024, krzyzanowski_machine_2024, yuan2025qme14scomprehensiveefficientspectral, Cai, xu_pretrained_2025, falletta2025unified, bhatia2025}.
However, current MLIPs capable of simulating IR spectra remain limited to narrow chemical domains and have not yet been systematically tested against molecular dynamics-based IR spectra predictions \cite{zou_deep_2023, zhang_universal_2023, Pracht_2024, MACE_off_2025, kabylda_molecular_2025, xu_pretrained_2025}. 
This narrow chemical coverage underscores the challenge of extending MLIPs beyond familiar molecular families and motivates the development of a universal model capable of spanning a broad chemical space. Yet, as the range of applicability grows, so does the risk of encountering chemistries that were not represented during training, making it difficult to know when predictions can be trusted. 
Therefore, incorporating uncertainty quantification into MLIPs would provide a direct measure of confidence, enabling users to identify trustworthy predictions and flag regions where the model may be less reliable. Still, to date, no universal MLIP has provided a convenient way to estimate uncertainty in IR spectral predictions, leaving a gap in the development of robust and transferable models capable of handling diverse molecular families and enabling fast, accurate simulations at scale. Such capabilities are crucial for supporting molecular identification from experimental data.

In this work, we introduce \MIR, an uncertainty-aware foundation model designed to deliver accurate, reliable, and efficient IR spectral predictions for molecules across a wide chemical space. \MIR{} is trained on a large-scale dataset of approximately 16 million DFT-calculated geometries from the QCML database \cite{ganscha_qcml_2025}, covering around 80 chemical elements and a diverse set of molecular classes. The model builds on the equivariant message-passing neural network \texttt{MACE} \cite{Batatia2022Design, MACE}, which has demonstrated strong performance in both IR property prediction \cite{bhatia2025} and general-purpose MLIP development \cite{MACE_off_2025, batatia2024foundationmodelatomisticmaterials}.
By formulating \MIR{} as an ensemble of independently trained models, we seek to systematically quantify predictive uncertainty and assess model reliability beyond the training domain.
To further support flexible and practical simulation workflows, we provide separate ensembles trained with different treatments of long-range dispersion interactions, including many-body dispersion (MBD) \cite{hermann2020density}, DFT-D4 \cite{caldeweyher2019generally}, and variants without explicit dispersion corrections. This allows users to select a level of theory that best matches their downstream applications.

We target assessing the robustness of \MIR{} by comparing predicted harmonic IR spectra and molecular dynamics-based IR spectra, which capture temperature-dependent anharmonic effects, with reference DFT calculations and experimental data. We aim to capture NQEs in the spectra through PIMD simulations, further enhancing the predictive fidelity of \MIR{}.
Testing across diverse molecular families allows us to evaluate the model’s transferability.
By doing so, we seek to provide the community with an uncertainty-aware foundation model for high-fidelity spectroscopic applications, usable as-is or adaptable to specific molecules through fine-tuning.

\section{Results}\label{results}

\begin{figure}[t!]
	\centering 
    \includegraphics[width=1.0\textwidth]{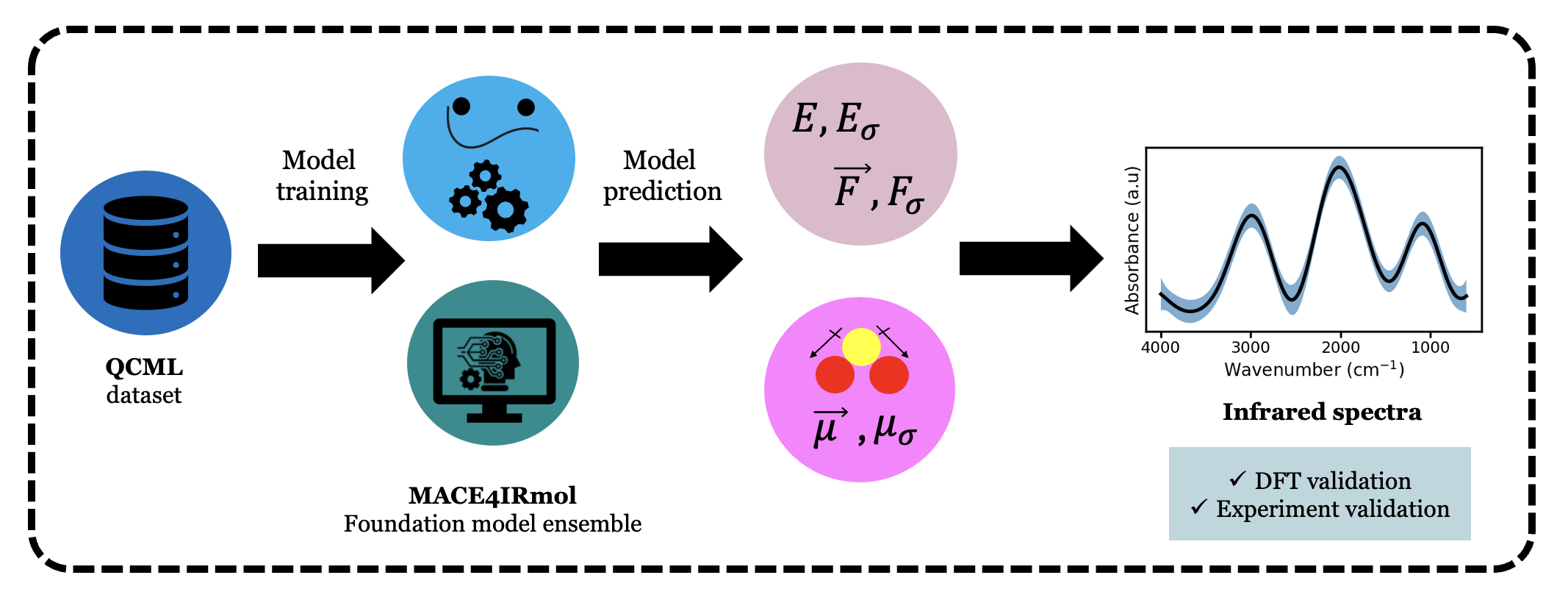}	
	\caption{
    \textbf{Illustration of the \MIR{} model development, ensemble-based uncertainty estimation, and performance evaluation in IR spectroscopy.}
    A chemically diverse QCML dataset is used to train an ensemble of \MIR{} models that predict energies ($E$), forces ($\Vec{F}$), and dipole moments ($\Vec{\mu}$). 
    Ensemble inference yields mean predictions together with uncertainty-aware quantities 
    $(\mathbf{E}_{\sigma}, \mathbf{F}_{\sigma}, \boldsymbol{\mu}_{\sigma})$, 
    where the subscript $\sigma$ denotes the ensemble-derived predictive uncertainty associated with each observable. 
    These uncertainties are propagated through the harmonic or MD-based IR simulation pipelines, producing spectra with uncertainty estimates. }
	\label{Combined_workflow}%
\end{figure}

We begin by providing an overview of the \MIR{} training workflow and evaluations of the trained foundation model, as illustrated in 
Figure~\ref{Combined_workflow}. Training starts with a curated subset of the QCML dataset~\cite{ganscha_qcml_2025}, which provides DFT-level energies, forces, and dipole moments. The QCML dataset also contains additional dispersion corrections for energies and forces on the  many-body dispersion (MBD) \cite{hermann2020density} and DFT-D4 \cite{caldeweyher2019generally} level.
The \MIR{} framework is implemented as a suite of equivariant \MACE{} models, comprising two distinct components: an interatomic potential for predicting energies and forces (MLIP), and a separate model for dipole moments. Together, these are referred to as the machine learning (ML) model throughout this work, and are trained to predict the reference properties.
For each component, we train an ensemble, which provides robust uncertainty estimation alongside the predictions. The \MIR{} MLIP  suite also includes variants trained on pure DFT data as well as those with added MBD and DFT-D4 corrections.
In the final part, we compare the \MIR{} outputs to reference DFT data from multiple databases. \MIR{} is then used to simulate IR spectra at three levels of theory: (i) the harmonic approximation, (ii) molecular dynamics (MD) simulations to capture anharmonic and temperature-dependent effects, and (iii) PIMD simulations accounting for NQEs, enabling more accurate IR spectral predictions for molecules where quantum nuclear motion is significant.
These predictions, together with their associated uncertainty estimates, are then validated against DFT simulations and experimental spectra for selected molecules. Finally, we assess the model’s transferability by applying it to molecules with diverse chemical compositions and structural complexities. Unless otherwise stated, all results presented in this work are obtained using the ensemble trained without any explicit dispersion correction.

In the following, we describe each component of the training workflow and present the results of \MIR{} predictions, including comparisons with DFT data, analysis of uncertainty estimates against true errors, and IR spectral simulations benchmarked against DFT reference and experimental data.

\textbf{Training data overview:}
We use a selected subset of the QCML dataset~\cite{ganscha_qcml_2025}. The original QCML dataset was constructed by generating conformations from 17.2 million unique chemical graphs, sampled across a wide temperature range (0–1000 K). Initial properties were computed using semi-empirical methods, and a representative set of 33.5 million conformers was subsequently recomputed using DFT to obtain accurate energies, atomic forces, and dipole moments~\cite{ganscha_qcml_2025}. From this pool, we retained only neutral molecules with singlet spin multiplicity (\(\sim\)16 million molecules) and randomly selected 10 million diverse entries spanning \(\sim\)80 elements of the periodic table. For different models in the \MIR{} ensemble, independent random splits were used, thus the ensemble covers  all \(\sim\)16 million molecules. 
The same splits were used for models trained on pure DFT data and those including MBD or DFT-D4 dispersion corrections. For more details on the filtering criteria and data selection, we refer to the Methods section. 

\begin{figure}[t!]
	\centering 
    \includegraphics[width=1.0\textwidth]{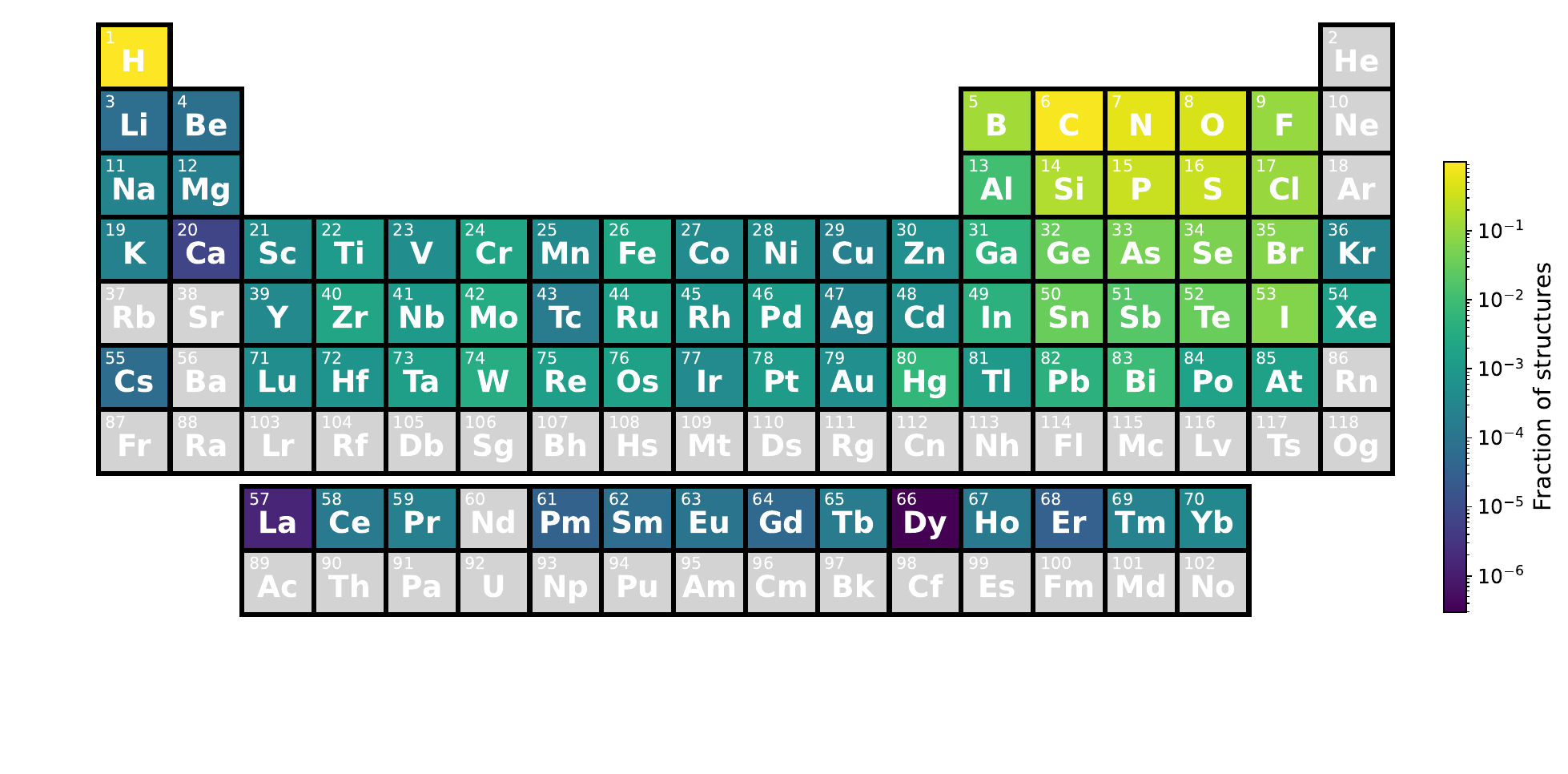}	
	\caption{Elemental distribution in the filtered 10M QCML dataset used to train the first ensemble model, highlighting chemical diversity across the periodic table. The color indicates which fraction of structures contains a given element (greyed out entries are not contained in any structure). The majority of structures are composed of H, C, N, O, P, and S, while nearly all elements with atomic number $Z < 86$ are represented in at least a few structures.
    }
	\label{Dataset_diversity}%
\end{figure}

The molecules are structurally and chemically diverse and include organic, inorganic, metal complexes, and biologically relevant species~\cite{ganscha_qcml_2025}. This broad coverage is illustrated in Fig.~\ref{Dataset_diversity}, which shows the distribution of elements represented in the dataset used to train the first model of the \MIR{} ensemble. Lighter elements such as H, C, N, O, S, and P are most prevalent, reflecting their ubiquity in organic and biochemical molecules. QCML also features molecules containing a wide range of heavier main-group and transition-metal elements. Similar elemental distribution analyses were performed for the remaining ensemble models and show consistent trends; these results are provided in the Supplementary Information (Figures~S1 and S2). This broad elemental diversity enables \MIR{} to learn different chemical patterns across varied bonding environments and coordination motifs, facilitating generalization to complex molecular systems encountered in catalysis, materials science, and environmental applications.

\begin{table}
\centering
\caption{Mean absolute errors (MAEs) for energy and force predictions on the 100k QCML-large test set, evaluated across varying training set sizes and model architectures. ``Small'', ``Medium'', and ``Large'' denote different \texttt{MACE-EF} model architectures, defined by the number of channels and maximum angular momentum as summarized in Table~\ref{table_mace_architecture}. The best-performing model, used in all inference tasks, is highlighted in bold.}
\label{table_training_curve}
\small
\begin{tabularx}{\textwidth}{>{\centering\arraybackslash}X >{\centering\arraybackslash}X c c}
\toprule
\textbf{Training set size} &
\textbf{Model configuration} &
\makecell[c]{\textbf{Energy} \\ (meV/atom)} &
\makecell[c]{\textbf{Force} \\ (meV/Å)} \\
\midrule

\multicolumn{4}{c}{\textbf{Training set size scaling — Small models (128 channels, $L{=}1$)}} \\
\midrule
50k  & Small / float32  & 10.7 & 98.9 \\
100k & Small / float32  & 14.1 & 96.6 \\
500k & Small / float32  & 8.1  & 75.5 \\
1M   & Small / float32  & 5.1  & 62.7 \\
3M   & Small / float32  & 3.8  & 50.1 \\
7M   & Small / float32  & 3.4  & 46.2 \\
10M  & Small / float32  & 3.3  & 44.4 \\
10M  & Small / float64  & 3.1  & 43.9 \\

\midrule
\multicolumn{4}{c}{\textbf{10M architecture comparison — Medium models (196 channels, $L{=}1$)}} \\
\midrule
10M & Medium / float32 & 2.5 & 32.3 \\
10M & Medium / float64 & 2.3 & 32.1 \\

\midrule
\multicolumn{4}{c}{\textbf{10M architecture comparison — Large models (256 channels, $L{=}2$)}} \\
\midrule
10M & Large / float64  & \textbf{2.1} & \textbf{30.1} \\
\bottomrule
\end{tabularx}
\end{table}

\textbf{ML model overview:} For \MIR, we employ the \MACE{} architecture~\cite{Batatia2022Design, MACE}, an equivariant message passing neural network designed to capture complex atomic interactions while respecting the symmetries of physical systems. Specifically, we train two separate models: \texttt{MACE-EF}, which predicts total energies and atomic forces in the role of a standard MLIP, and \texttt{MACE-D} for predicting dipole moments. For each model type, we train an ensemble of three independently initialized models to improve robustness and uncertainty estimation. Together, these ensembles constitute the \MIR{} foundation model and are released under the name \texttt{MACE4IRmol}. 
The hyperparameters of these models can be found in the Methods section.

\begin{table}
\centering
\caption{Mean absolute errors (MAEs) for dipole moment predictions on the 100k QCML-large test set. Results are reported for varying training set sizes (top section) and for different \texttt{MACE-D} model architectures trained on 10M molecular structures (bottom section). The model highlighted in bold, was used for all inference tasks.}
\label{table_dipole_training_curve}
\small
\begin{tabularx}{\textwidth}{p{2.7cm} >{\centering\arraybackslash}X c}
\toprule
\textbf{Training set size} &
\textbf{Model configuration} &
\makecell[c]{\textbf{Dipole moment} \\ (meÅ)} \\
\midrule
\multicolumn{3}{c}{\textbf{Training set size scaling}} \\
\midrule
50k  &                                          & 41.7 \\
100k &                                          & 36.9 \\
500k &                                          & 29.6 \\
1M   & Small / 32 channels /  $L{=}1$ / float32 & 26.3 \\
3M   &                                          & 25.3 \\
7M   &                                          & 23.5 \\
10M  &                                          & 23.3 \\
\midrule
\multicolumn{3}{c}{\textbf{10M dipole model architecture comparison}} \\
\midrule
10M & Small / 16 channels /  $L{=}1$ /  float64            & 28.5 \\
10M & Small / 32 channels /  $L{=}1$ /  float64            & 23.3 \\
10M & Medium / 64 channels /  $L{=}1$ /  float32           & 21.1 \\
10M & Medium / 92 channels /  $L{=}1$ /  float32  & 21.5 \\
10M & Medium / 92 channels /  $L{=}2$ /  float32  & 21.2 \\
10M & Large / 128 channels /  $L{=}1$ /  float32  & \textbf{20.9} \\
10M & Large / 128 channels /  $L{=}2$ /  float32  & 20.8 \\
10M & Large / 256 channels /  $L{=}1$ /  float32  & 20.9 \\
\bottomrule
\end{tabularx}
\end{table}

\textbf{Training and performance evaluation of \MIR{} ensemble models:}
To assess how model performance scales with training set size and to establish a robust ensemble framework, we implemented a hierarchical training strategy. Starting from a filtered pool of approximately 16 million neutral, singlet molecular structures from the QCML dataset, we randomly sampled three independent subsets of 10 million structures each. These subsets were used to train independent models, enabling ensemble construction while keeping the validation and test sets fixed. For each 10M subset, we further generated training splits containing 50k, 100k, 500k, 1M, 3M, 7M, and the full 10M structures. A fixed validation set of 100,000 structures and an independent test set of 100,000 structures were used consistently across all experiments; the test set is referred to as the "QCML-large" test set.

\begin{figure}[t!]
	\centering 
    \includegraphics[width=0.6\textwidth]{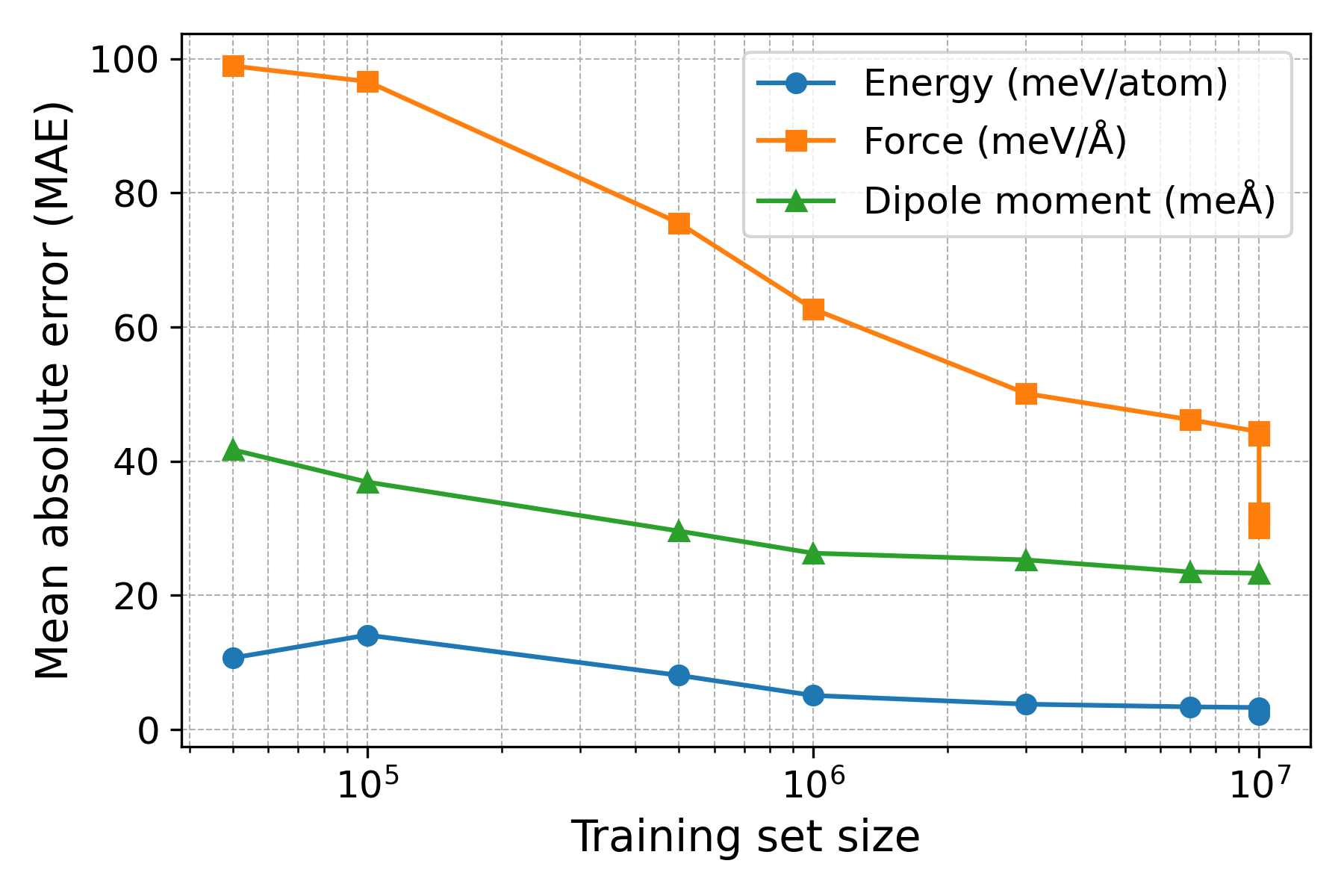}	
	\caption{Mean absolute error (MAE) of energy, force and dipole moment predictions on the test set for the first ensemble model trained on each given number of examples from the training data. }
	\label{learning_curve}%
\end{figure}

For all training subsets, \texttt{MACE-EF} and \texttt{MACE-D} models were trained independently to optimize each for its respective target property. Model performance was evaluated using the mean absolute error (MAE), with learning curves shown in Fig.~\ref{learning_curve} and quantitative results summarized in Table~\ref{table_training_curve} and Table~\ref{table_dipole_training_curve}. In all cases, prediction errors decrease systematically with increasing training set size.

\textbf{Energy and force prediction with \texttt{MACE-EF}:}
For energy and force prediction, the best performance is achieved using the Large \texttt{MACE-EF} architecture trained on the full 10M dataset with double precision (float64), yielding MAEs of 2.1 meV/atom for energies and 30 meV/Å for forces (see Table~\ref{table_training_curve}). Owing to its superior accuracy and numerical stability, this configuration is used throughout the work for energy–force predictions and for ensemble construction.

The ensemble of three models, allowing for uncertainty quantification, was created only for the 10M training set and Large, float64 \texttt{MACE-EF} model. Simultaneously, we have prepared three ensembles, corresponding to different treatments of long-range dispersion interactions. Specifically, separate ensembles were constructed for models without an explicit dispersion correction (PBE0), with many-body dispersion (PBE0+MBD), and with the DFT-D4 dispersion correction (PBE0+D4). For each interaction type, three models were trained using identical hyperparameters, resulting in a total of nine \texttt{MACE-EF} models across all interaction settings. Ensemble-averaged errors and standard deviations are reported in Table~\ref{table_maceef_ensemble}, demonstrating both high predictive accuracy and low variance across ensemble members.

\begin{table}
\centering
\caption{Performance of the \texttt{MACE-EF} ensembles trained on 10M molecular structures using the Large, float64 architecture.
Each ensemble consists of three independently trained models per DFT reference.
Mean and standard deviation (std) are reported across ensemble members.}
\label{table_maceef_ensemble}
\small
\begin{tabularx}{\textwidth}{>{\centering\arraybackslash}X c c}
\toprule
\multicolumn{3}{c}{\textbf{Model configuration: 10M Large (float64)}} \\
\midrule
\textbf{DFT reference} &
\makecell[c]{\textbf{Energy} \\ (meV/atom)} &
\makecell[c]{\textbf{Force} \\ (meV/Å)} \\
\midrule
PBE0       & 2.1 & 30.1 \\
PBE0       & 2.0 & 28.6 \\
PBE0       & 2.1 & 28.8 \\
\textbf{PBE0 (mean ± std)} & \textbf{2.07 ± 0.05} & \textbf{29.17 ± 0.66} \\
\midrule
PBE0 + D4  & 1.9 & 27.1 \\
PBE0 + D4  & 2.0 & 28.5 \\
PBE0 + D4  & 2.1 & 28.7 \\
\textbf{PBE0 + D4 (mean ± std)} & \textbf{2.00 ± 0.08} & \textbf{28.10 ± 0.71} \\
\midrule
PBE0 + MBD & 1.9 & 26.9 \\
PBE0 + MBD & 2.0 & 28.5 \\
PBE0 + MBD & 2.1 & 28.8 \\
\textbf{PBE0 + MBD (mean ± std)} & \textbf{2.00 ± 0.08} & \textbf{28.07 ± 0.83} \\
\bottomrule
\end{tabularx}
\end{table}

\begin{table}
\centering
\caption{Ensemble performance of selected \texttt{MACE-D} models trained on 10M molecular structures.
Each ensemble consists of three independently trained models with identical architectures.}
\label{table_maced_dipole_ensemble}
\small
\begin{tabularx}{\textwidth}{>{\centering\arraybackslash}X c}
\toprule
\textbf{Model configuration} &
\makecell[c]{\textbf{Dipole moment} \\ (meÅ)} \\
\midrule
10M Small (64 channels, $L{=}1$, float32) & 21.1 \\
                                 & 21.1 \\
                                 & 22.4 \\
\textbf{Mean ± std}              & \textbf{21.53 ± 0.75} \\
\midrule
10M Small (92 channels, $L{=}1$, float32) & 21.5 \\
                                          & 21.8 \\
                                          & 21.6 \\
\textbf{Mean ± std}                        & \textbf{21.63 ± 0.15} \\
\midrule
10M Small (128 channels, $L{=}1$, float32) & 20.9 \\
                                                     & 20.7 \\
                                                     & 20.9 \\
\textbf{Mean ± std}                                 & \textbf{20.83 ± 0.12} \\
\bottomrule
\end{tabularx}
\end{table}

\textbf{Dipole moment prediction with \texttt{MACE-D}:}
For dipole moment prediction, we explored multiple \texttt{MACE-D} architectures at the 10M scale, varying channel width, angular momentum cutoff, and numerical precision (see Table~\ref{table_dipole_training_curve}). Interestingly, double precision (float64) offered no accuracy advantage over single precision (float32), and we therefore adopted single precision for the Medium and Large models. Compared to the previous models in Table~\ref{table_dipole_training_curve}, the last two required prohibitively long training times (see Table S1 in the Supplementary Information) without improving accuracy, and were therefore excluded from further consideration.

In order to distinguish between the closely performing Medium and Large models, we carried out additional evaluations not only on the QCML-large test set but also on chemically distinct external benchmarks. The candidate dipole models were assessed on: (a) a representative organic subset of QM7-x \cite{donchev2021quantum} (300 structures containing H, C, N, and O), (b) a transition-metal-focused subset derived from tmQM \cite{balcells_tmqm_2020} (50 metal coordination complexes), and (c) three trajectories extracted from AIMD simulations of representative molecules (10,000 randomly selected configurations each) to probe stability along dynamical pathways. Details of all test sets are provided in the Methods section. A comprehensive comparison of dipole model performance across these benchmarks is reported in Table S2 of the Supplementary Information.

Based on this combined evaluation, we selected the 128-channel, $L{=}1$ \texttt{MACE-D} architecture as the primary model for ensemble construction and all dipole-related prediction tasks, balancing accuracy and computational efficiency. To quantify uncertainty, ensembles of three independently trained models with this architecture were constructed. In addition, ensembles of computationally cheaper dipole models (64- and 92-channel variants) were also trained and analyzed, as summarized in Table~\ref{table_maced_dipole_ensemble}, and can be used in selected benchmarks where higher throughput is required.

\textbf{Model training times:} The best-performing 10M \texttt{MACE-EF} model (Large, float64) was trained in approximately three days using 24 AMD MI250x GPU, whereas the corresponding 10M \texttt{MACE-D} model required approximately nine days of training on a single AMD MI250x GPU.

\textbf{Evaluation on external test sets:}
We next evaluate the performance and transferability of the resulting \MIR{} ensemble models on the external QM7-x and tmQM test subsets, which probe chemically distinct regimes not fully represented in the QCML training data. Quantitative results for these evaluations are summarized in Table~\ref{table_external_tests}. Overall, \MIR{} maintains high accuracy on the QM7-x subset, which consists of organic molecules with light elements. In contrast, prediction errors are substantially larger for the tmQM subset, which contains transition-metal complexes.

\begin{table}[t]
\centering
\caption{Mean absolute errors (MAEs) of \MIR{} ensemble predictions on external test sets probing chemically distinct regimes. Reported metrics include energy per atom, atomic forces, and dipole moment components.}
\label{table_external_tests}
\small
\begin{tabularx}{\textwidth}{l c c c}
\toprule
\textbf{Dataset} &
\makecell[c]{\textbf{Energy MAE} \\ (meV/atom)} &
\makecell[c]{\textbf{Force MAE} \\ (meV/\AA)} &
\makecell[c]{\textbf{Dipole MAE} \\ (e\AA)} \\
\midrule
QM7-x (organic molecules)      & 2.39   & 36.91  & 0.037 \\
tmQM (transition-metal complexes) & 144.97 & 433.83 & 0.292 \\
\bottomrule
\end{tabularx}
\end{table}

\textbf{Uncertainty quantification through \MIR{} ensemble:}
To evaluate model uncertainty, we computed ensemble predictions on three new and distinct test sets. These include the subset of QCML-large test set, called QCML-small, that includes 672 molecules spanning 76 elements. QCML-small was constructed to sample chemically diverse structures containing rare main-group and transition-metal elements, as well as the QM7-x and tmQM subsets, representing conventional organic molecules and transition-metal-containing systems, respectively. More details on the construction of these test sets are provided in the Methods section. Uncertainty estimates were computed for three target properties: total energies, per-atom forces, and dipole moments.

\begin{figure}[t!]
\centering
\includegraphics[width=1.0\textwidth]{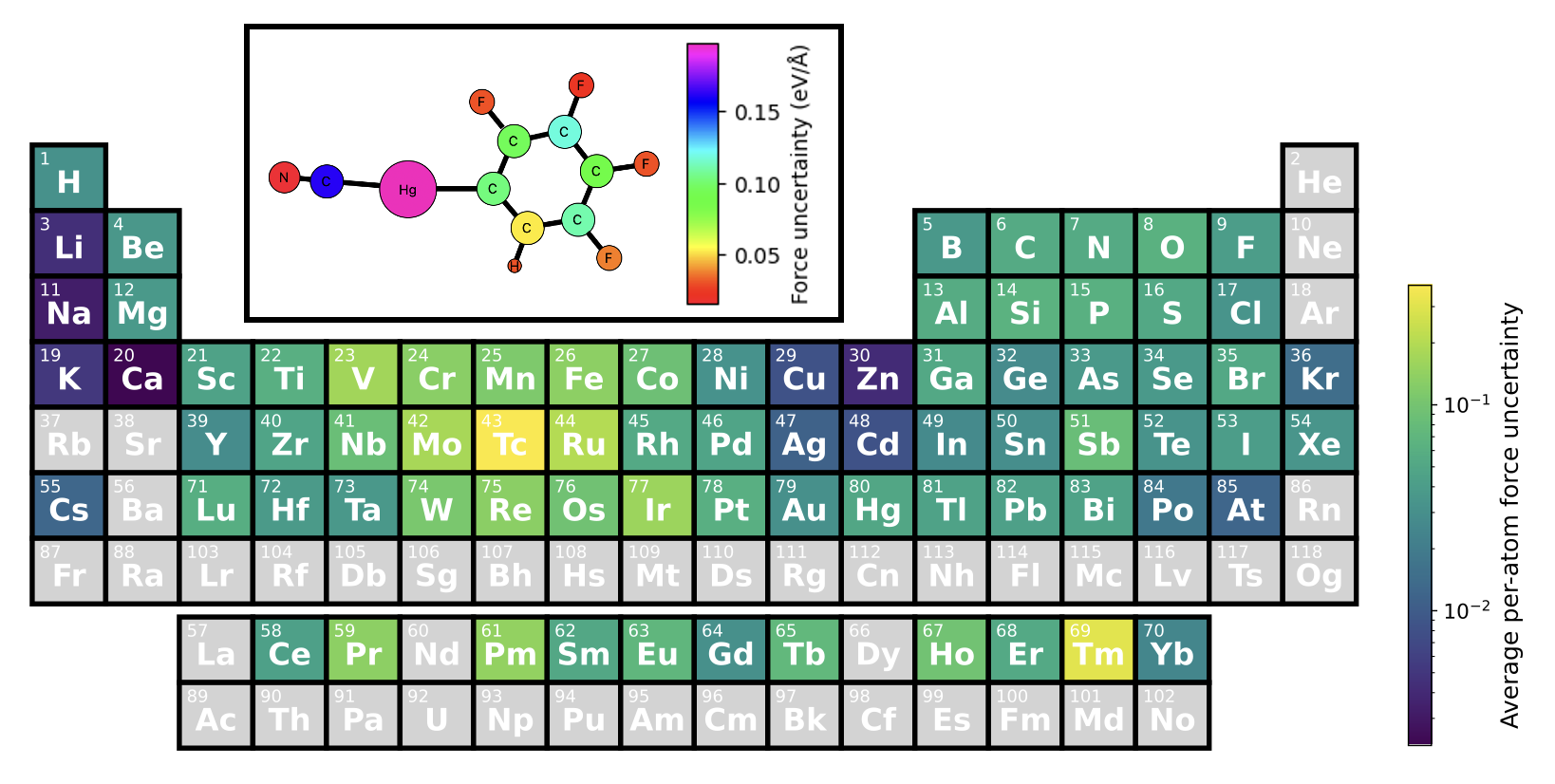}
\caption{
Element-resolved analysis of ensemble-based uncertainty in per-atom force predictions for the QCML-small test set. The heatmap shows the distribution of force prediction uncertainties across chemical elements, highlighting systematic variations in model confidence as a function of atomic species in chemically diverse environments. The inset displays an example molecule in which atoms are colored by their predicted force uncertainty.
}
\label{fig:uncertainty_force_qcml}
\end{figure}

For per‑atom forces, we evaluated uncertainty at the level of individual atoms, as illustrated in the inset of Figure~\ref{fig:uncertainty_force_qcml}.
We provide element‑wise distributions for each of the three test sets, with Figure~\ref{fig:uncertainty_force_qcml} showing the element‑resolved distribution obtained for the QCML‑small test set.
A clear dependence of predictive uncertainty on atomic species is observed. Light main-group elements such as H, C, N, and O exhibit consistently low force uncertainties, reflecting stable and well-constrained predictions across diverse chemical environments. Slightly higher but still moderate uncertainties are observed for second- and third-row main-group elements, including S, P, and halogens.

In contrast, substantially elevated uncertainties are observed for heavier main-group elements and transition metals. Elements such as Ti, V, Cr, Mn, Fe, Co, and Ni display among the highest average force uncertainties, with similarly pronounced values extending across late transition metals and lanthanides. These trends persist across multiple coordination environments present in the QCML-small test set and indicate systematic variations in ensemble spread as a function of elemental identity.

Overall, the heatmap highlights that force prediction uncertainty increases with chemical complexity and atomic number, particularly for elements associated with diverse bonding patterns and coordination geometries.
Corresponding element-wise uncertainty distributions for the QM7-x and tmQM test subsets are provided in the Supplementary Information (Figures~S3 and S4) and show consistent behavior within their respective chemical domains. 

For total energy and dipole moment, which are global properties, we report the element-wise uncertainty distribution only for the QCML-small test set which was explicitly constructed to contain rare elements in isolation.
Similar trends are observed for total energy (Figure~S5) and dipole moment (Figure~S6) as seen for per-atom forces, with uncertainties increasing for heavier main-group elements, transition metals, and lanthanides. However, a comparison between the two global properties shows that the overall uncertainty in total energy is generally higher than that for dipole moments, particularly for transition metal elements and lanthanides. The full distributions of predicted energy and dipole moment uncertainties across all test sets, illustrating the overall spread of values, are provided in the Supplementary Information (Figures~S7–S9).

\textbf{Harmonic IR spectra predictions:}

\begin{figure}[t!]
\centering
\includegraphics[width=1.0\textwidth]{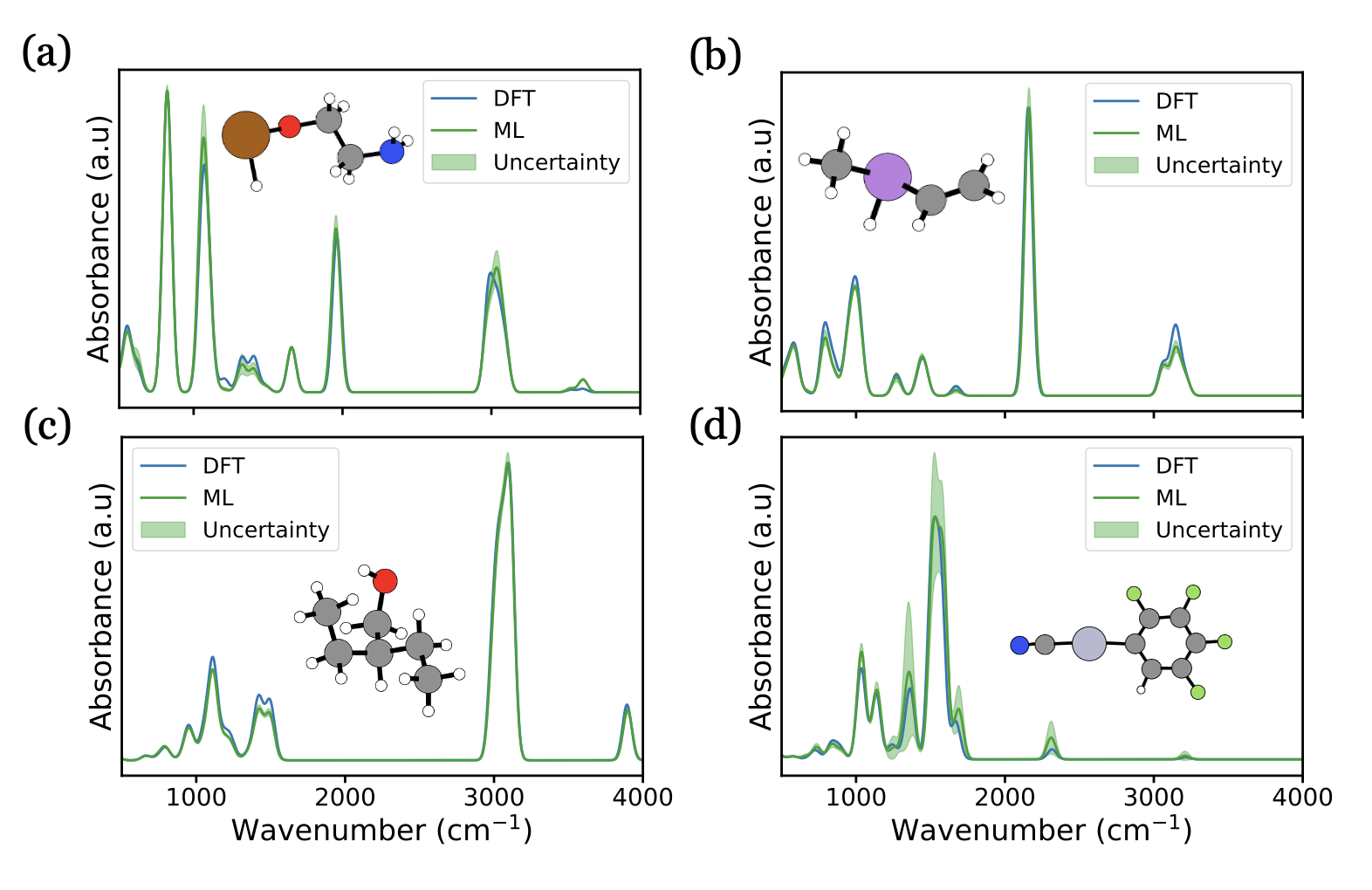}
\caption{
Comparison of DFT-calculated reference harmonic IR spectra with the ML-predicted ensemble (\MIR{}) average spectra for a representative set of molecules, with ensemble uncertainty also indicated.
}
\label{Harmonic_spectra}
\end{figure}

After validating that \MIR{} can accurately predict energies, forces, and dipole moments, we proceed to the prediction of IR spectra, starting with the harmonic approximation. For each molecule, harmonic IR spectra were predicted using the \MIR{} ensemble. Harmonic frequencies and intensities were computed independently with each ensemble member, and the final ML prediction was obtained by averaging over the ensemble. The standard deviation across ensemble members provides a direct estimate of predictive uncertainty for both frequencies and intensities. These ensemble-averaged predictions were then compared against DFT reference spectra to compute errors and assess model reliability.

Representative harmonic spectra for four chemically distinct molecules are shown in Fig.~\ref{Harmonic_spectra}, illustrating the close agreement between predicted and reference peak positions and relative intensities. In order to systematically assess how accurately the spectra are predicted, we benchmarked \MIR{} across diverse chemical spaces, using the three external test sets introduced earlier: the QCML-small test set (672 molecules), the QM7-x subset (300 molecules), and the tmQM subset (50 molecules). Together, these datasets probe rare-element chemistry, conventional organic molecules, and transition-metal-containing systems, respectively. Details of the test set preparation, geometry optimization protocol, and filtering criteria are provided in the Methods section.
More detailed statistics on the filtered dataset can be found in Section~1.0 of the Supplementary Information.

\begin{table}[t]
\centering
\caption{Mean absolute errors (MAEs) for harmonic vibrational frequencies and IR intensities across three test datasets.
Errors are reported overall and separately for low- ($\leq$1000~cm$^{-1}$) and high-frequency ($>$1000~cm$^{-1}$) spectral regions.}
\label{table_harmonic_ir}
\small
\begin{tabularx}{\textwidth}{>{\raggedright\arraybackslash}X l c c}
\toprule
\textbf{Dataset} &
\textbf{Spectral regime} &
\makecell[c]{\textbf{Frequency MAE} \\ (cm$^{-1}$)} &
\makecell[c]{\textbf{Intensity MAE} \\ ((D/\AA)$^2$ amu$^{-1}$)} \\
\midrule
\multirow{3}{*}{\makecell[l]{QCML-small test set \\ ($\sim$76 elements)}}
& Overall                     & 24.65 & 0.94 \\
& $\leq$1000~cm$^{-1}$         & 28.23   & 0.52  \\
& $>$1000~cm$^{-1}$            & 21.14   & 1.44  \\
\midrule
\multirow{3}{*}{\makecell[l]{QM7-x subset \\ (H, C, N, O)}}
& Overall                     & 2.74  & 0.11 \\
& $\leq$1000~cm$^{-1}$         & 3.21   & 0.06  \\
& $>$1000~cm$^{-1}$            & 2.45   & 0.14 \\
\midrule
\multirow{3}{*}{\makecell[l]{tmQM subset \\ (transition metals)}}
& Overall                     & 25.25 & 0.84 \\
& $\leq$1000~cm$^{-1}$         & 28.40 & 0.38 \\
& $>$1000~cm$^{-1}$            & 22.65 & 1.57 \\
\bottomrule
\end{tabularx}
\end{table}

For each dataset, we report MAEs in vibrational frequencies and IR intensities and summarize in Table~\ref{table_harmonic_ir}. The MAE values indicate that \MIR{} achieves high accuracy in predicting harmonic vibrational frequencies and IR intensities across all three test sets. For the organic QM7-x subset, the model exhibits the lowest errors, with an overall frequency MAE of 2.74~cm$^{-1}$ and intensity MAE of 0.11~(D/\AA)$^2$~amu$^{-1}$. In contrast, the QCML-small and tmQM test subsets show higher overall frequency errors, 24.65~cm$^{-1}$ and 25.25~cm$^{-1}$ respectively, and correspondingly higher intensity MAEs, 0.94 and 0.84~(D/\AA)$^2$~amu$^{-1}$. 

When considering spectral regions separately, low-frequency modes ($\leq$1000~cm$^{-1}$) generally show slightly larger frequency errors than high-frequency modes ($>$1000~cm$^{-1}$) in QCML-small and tmQM, consistent with the increased anharmonicity and collective motion in these vibrations. Conversely, intensity predictions exhibit larger errors in the high-frequency region, particularly for QCML-small and tmQM. 

\begin{figure}[t!]
\centering
\includegraphics[width=1.0\textwidth]{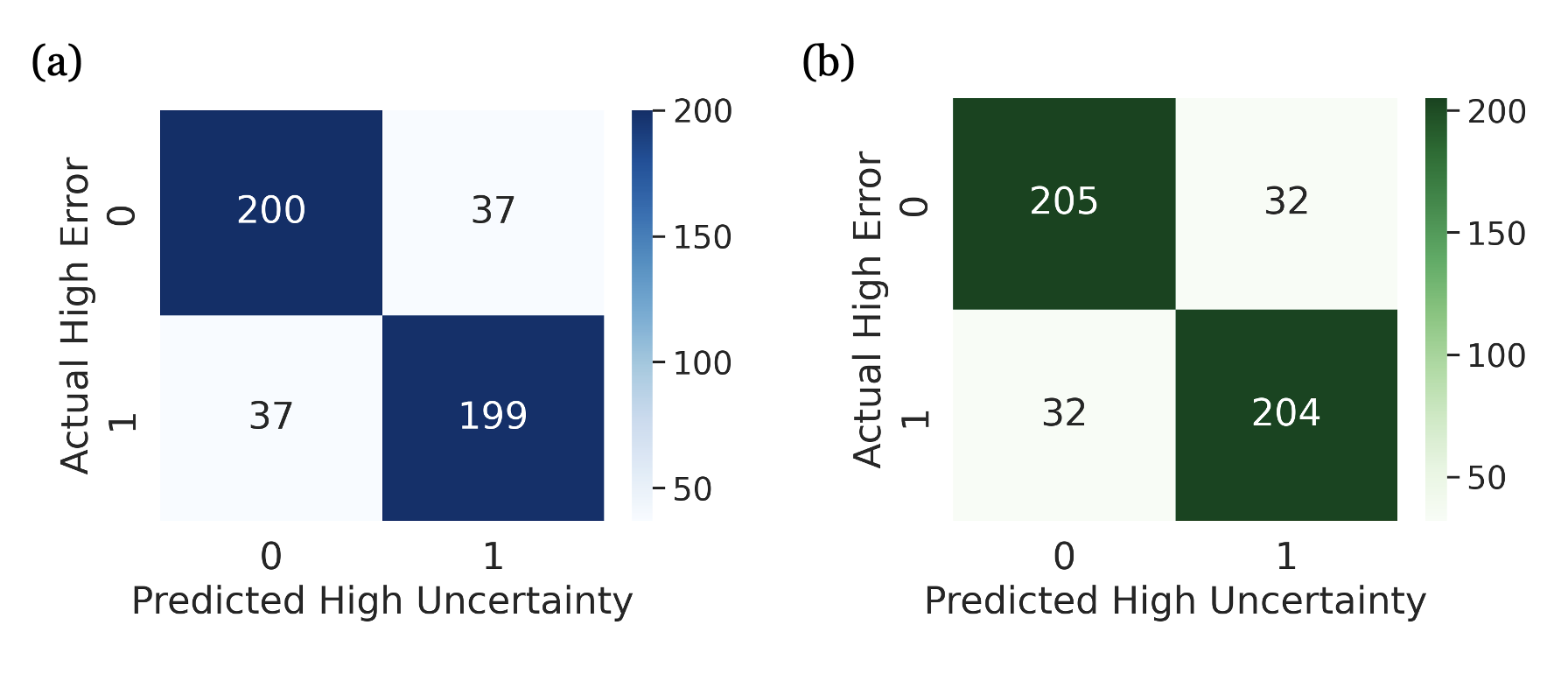}
\caption{
Confusion matrices for the QCML-small test set illustrating the relationship between ensemble-based uncertainty and prediction error for harmonic IR properties: (a) for vibrational frequencies and (b) for IR intensities. Predictions are categorized into low/high uncertainty and low/high error regimes using thresholds defined by the median values of the respective distributions. For vibrational frequencies, the median error is 9.4333~cm$^{-1}$ and the median uncertainty is 5.6933~cm$^{-1}$. For IR intensities, the median error is 0.3461~(D/\AA)$^2$~amu$^{-1}$ and the median uncertainty is 0.2235~(D/\AA)$^2$~amu$^{-1}$. The high diagonal numbers and low off-diagonal values demonstrates that elevated uncertainty consistently correlates with larger prediction errors.
}
\label{fig:qcml_confusion_matrix}
\end{figure}

Beyond average performance metrics, we examined the relationship between ensemble-predicted uncertainty and actual prediction error using two complementary approaches: the Pearson correlation between uncertainty and absolute error, and confusion matrices for both frequencies and intensities.
For the QCML-small harmonic benchmark subset, the Pearson correlation between frequency MAE and uncertainty is 0.79, while the correlation for IR intensity MAE is 0.96 (see Supplementary Figure~S10). The confusion matrices in Figure~\ref{fig:qcml_confusion_matrix} show that the ensemble achieves an accuracy of 84.36\% for frequencies and 86.47\% for IR intensities.

Finally, for the small fraction of high-error cases (MAE $>$ 200~cm$^{-1}$), the Pearson correlation between frequency MAE and ensemble-predicted uncertainty is 0.98, indicating strong positive relationship between uncertainty and error.
Similar trends are observed for the QM7-x and tmQM subsets (see Supplementary Figures~S11–S14).

Furthermore, example molecules exhibiting particularly large prediction errors, as well as cases for which the DFT reference calculations did not converge, are highlighted in the Supplementary Information Figure~S15.

In terms of computational efficiency, DFT-based calculations require $\sim$180 CPU hours for a 14-atom molecule, the ML model predicts the spectra in 10 seconds on a single GPU. 

\textbf{Molecular dynamics-based IR spectra predictions:}
We computed MD-based IR spectra from dipole moment trajectories at 300~K using both classical MD (named ML-MD hereafter) and quantum effects with thermostatted ring-polymer PIMD with generalized Langevin equation thermostats (TRPMD-GLE) \cite{rossi2014remove, rossi2018fine}, which we denote as ML-PIMD for simplicity. We selected TRPMD-GLE based on the criteria mentioned in the Methods section.
Further computational details are also provided in the Methods section. 

\begin{figure}[t!]
    \centering 
    \includegraphics[width=1.0\textwidth]{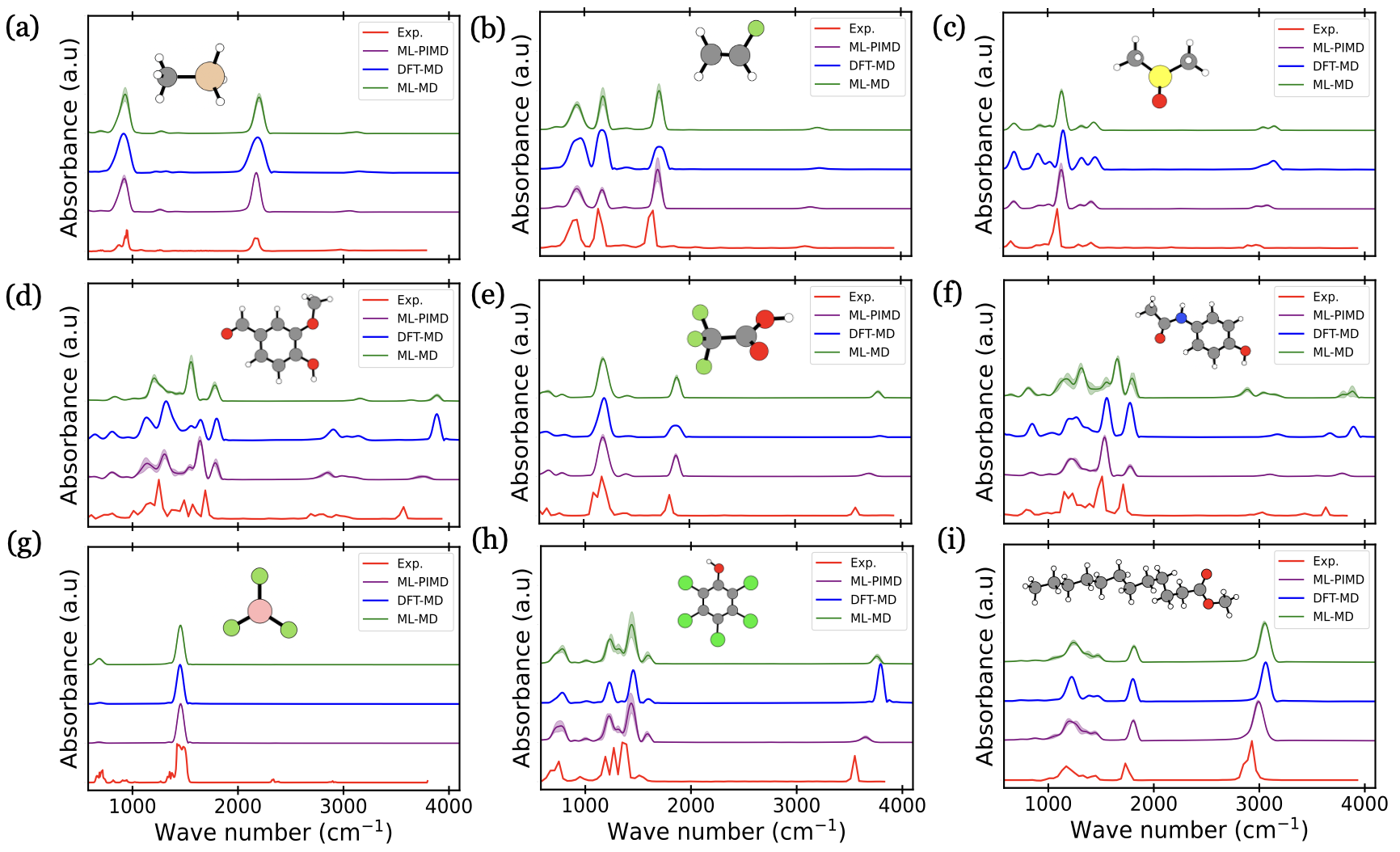}	
    \caption{
    Comparison of ML-predicted (\MIR) IR spectra with DFT-MD reference and experimental spectra from the NIST databasefor a diverse set of molecules at 300K, 
    All ML predictions, ML-MD and ML-PIMD, show the ensemble average, with uncertainties indicated:  
    (a) Methylsilane,  
    (b) Fluoroethene,  
    (c) Dimethyl sulfoxide,  
    (d) 4-Hydroxy-3-methoxybenzaldehyde (Vanillin),  
    (e) Trifluoroacetic acid,  
    (f) 4-Acetamidophenol (Paracetamol),  
    (g) Boron trifluoride,  
    (h) Pentachlorophenol,  
    (i) Methyl dodecanoate.  
    }
    \label{MLMD_spectra}
\end{figure}

To evaluate the quality of \MIR{}-predicted spectra, we selected 9 molecules from the NIST database \cite{Wallace2024} with available experimental IR spectra; none of these molecules are included in the QCML dataset. Figure~\ref{MLMD_spectra} compares the predicted spectra (ML-MD and ML-PIMD) with experimental measurements and DFT-based AIMD simulations (denoted as DFT-MD). For ML-based spectra, we averaged the results from three independent trajectories and also provided uncertainty. More details of the simulation protocol and uncertainty estimation are provided in the Methods section.

The ML spectra reproduce experimental peak positions and intensities well, particularly in the low-frequency region. In the high-frequency range (3000–4000~cm$^{-1}$), both DFT-MD and ML-MD spectra exhibit systematic blue shifts relative to experiment. ML-PIMD spectra show closer agreement in this region, with reduced peak shifts compared to both classical ML-MD and DFT-MD results.

Quantitative agreement with experiment is assessed using the Pearson correlation coefficient (PCC) and Wasserstein distance (WD), as commonly used in MD-based spectra comparison \cite{esch_quantitative_2021, bhatia2025}.  The average values across the 9-molecule test set are given in Table~\ref{table:avg_PCC_WD}.

\begin{table}[h!]
\centering
\caption{Average Pearson correlation coefficients (PCC) and Wasserstein distances (WD) between experimental (Exp) IR spectra and the different MD predictions across the 9-molecule set.}
\label{table:avg_PCC_WD}
\begin{tabular}{l c c}
\hline
Comparison & Avg. PCC & Avg. WD \\
\hline
Exp vs DFT-MD & 0.465 & 0.0322 \\
Exp vs ML-MD  & 0.519 & 0.0246 \\
Exp vs ML-PIMD & 0.576 & 0.0236 \\
\hline
\end{tabular}
\end{table}

Across the test set, the \MIR{} outperforms DFT-MD in both metrics. From the ML-based predictions ML-PIMD shows consistently stronger correlation with experiment than ML-MD, as indicated by the higher PCC and lower WD values in Table~\ref{table:avg_PCC_WD}. Individual results for each molecule, including the corresponding PCC and WD values, are summarized in Table~S3 in the Supplementary Information, where additional details on these quantitative metrics are also provided (see Section~2.0 in the Supplementary Information).

The computational cost of the MD simulations using \MIR{} scales linearly with system size, $\mathcal{O}(N_{atoms})$, in contrast to the quartic scaling, $\mathcal{O}(N_{electrons}^4)$, of DFT calculations with the hybrid PBE0 functional~\cite{PBE0}.
This difference in scaling is reflected in the computational efficiency, as DFT-based calculations require $\sim$9000 CPU hours for a molecule comprising 6 atoms, whereas the ML model predicts the AIMD-based spectra in only $\sim$2 hours and PIMD-based spectra in $\sim$22 hours on a single GPU.

\textbf{Assessing generalization over molecules with varying element occurrence:}
To evaluate how well \MIR{} generalizes, we examined its performance on molecules containing elements with varying frequencies of occurrence in the training data using ML-PIMD spectra. Instead of limiting the analysis to only the most underrepresented or overrepresented elements (see Figure~\ref{Dataset_diversity}), we curated a diverse set of molecules spanning a broad range of elemental frequencies. These molecules were obtained from the NIST database \cite{Wallace2024} and are distinct from the QCML dataset used for training, ensuring that this evaluation probes the model’s extrapolative capabilities. This setup allows assessment of how well the model captures vibrational features across molecules composed of both common and rare elements, while simultaneously monitoring prediction uncertainty.

\begin{figure}[t!]
\centering
\includegraphics[width=1.0\textwidth]{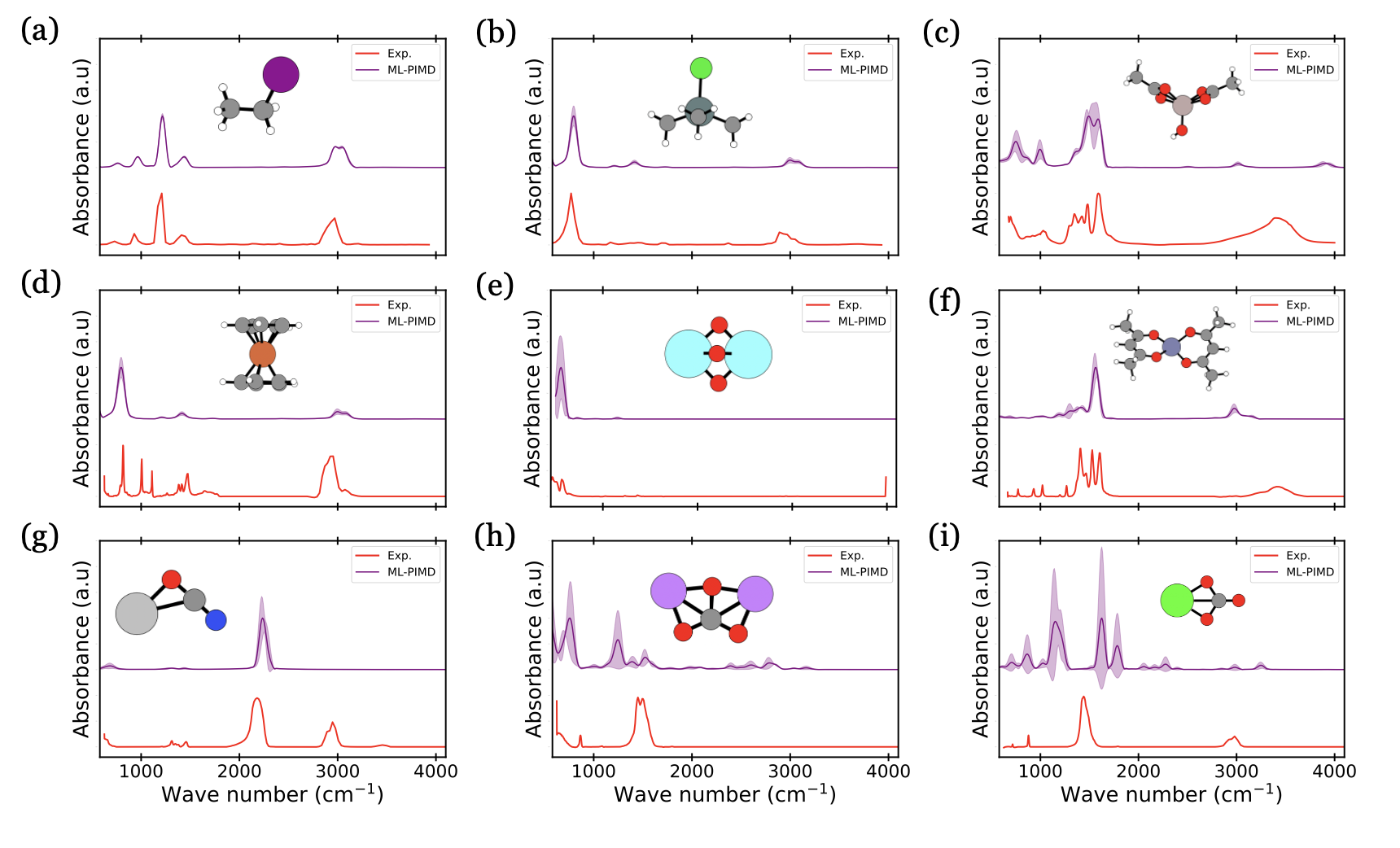}
\caption{
Comparison of ML-PIMD predicted IR spectra (\MIR{}) with experimental spectra from the NIST database for a representative set of molecules containing elements with varying frequencies in the training data at 300K. 
All ML predictions show the ensemble average, with uncertainties indicated.
(a) 1-Iodoethane
(b) Chlorotrimethylstannane
(c) Aluminum diacetate
(d) Ferrocene
(e) Yttrium oxide
(f) Bis(2,4-pentanedionato)zinc
(g) Silver cyanate
(h) Lithium carbonate
(i) Calcium carbonate.
The elemental representation in the training dataset gradually decreases from (a) to (i).
}
\label{Generability_MLMD}
\end{figure}

As shown in Figure~\ref{Generability_MLMD}, the molecules are arranged such that elemental abundance in the training dataset decreases progressively from (a) to (i). The comparison focuses on ML-PIMD spectra against experiment and includes ensemble-derived uncertainty estimates. For molecules (a) through (f), which primarily contain well-represented elements such as main-group species (H, C, N, O), halogens (Cl, I), and metals (Fe, Al, Sn, Zn, Y), the predicted spectra generally align closely with experiment and exhibit relatively low uncertainty. 
However, ferrocene (d), containing the transition metal Fe, represents an exception. Despite Fe being moderately represented in the training data, the ML-PIMD predictions for ferrocene show significantly poor agreement with experiment.
In contrast, molecules in the last row, (g) silver cyanate, (h) lithium carbonate, and (i) calcium carbonate, show more pronounced deviations from experiment. These systems contain elements (Ag, Li, Ca) that are sparsely represented in the training set, and the associated spectral uncertainty increases noticeably from (a) to (i). Quantitative metrics, including the PCC and WD (Table~S4), reveal a consistent decline in agreement with experiment across the series. 

To determine whether agreement between theory and experiment can already be predicted during the ML-PIMD simulations, we computed the mean and standard deviation of the predicted uncertainties for energy ($E_{\sigma}$), forces ($F_{\sigma}$), and dipole moments ($\mu_{\sigma}$) along the first ML-PIMD trajectory of each molecule. Table~S5 in Supplementary Information summarizes these values for the representative set of molecules shown in Figure~\ref{Generability_MLMD}.
As shown in Table~S5, molecules predominantly containing well-represented elements (e.g., 1-iodoethane, chlorotrimethylstannane, yttrium oxide) exhibit relatively low mean uncertainties across energy, forces, and dipole moments. In contrast, ferrocene (d), which contains a transition metal Fe, exhibits markedly higher force uncertainties. Molecules with less represented elements, such as silver cyanate, lithium carbonate, and calcium carbonate, correspondingly show increased predictive uncertainties, consistent with the decreasing coverage of these elements in the training data.

We further probe the high uncertainty observed for ferrocene. Table~S6 in the Supplementary Information lists the first occurrence of force uncertainty exceeding absolute thresholds (10, 20, 30, 50, 100), and the total simulation time. Notably, the force uncertainty exceeded 10 eV/\AA\ already at approximately 2 ps into the trajectory.

\section{Discussion}\label{sec4}

In this study, we have demonstrated that \MIR{} can serve as a foundation MLIP for molecular systems containing elements from nearly the entire periodic table, while also allowing for the inclusion of two dispersion‑correction treatments, DFT‑D4 and MBD. We have shown that it predicts energies, forces, and dipole moments with state‑of‑the‑art precision, not only for molecules from the QCML database (Table~\ref{table_maceef_ensemble} and ~\ref{table_maced_dipole_ensemble}) but also for molecular structures from external datasets (Table~\ref{table_external_tests}). Most importantly,  we have verified the ability of \MIR{} to predict IR spectra across three levels of accuracy: the harmonic approximation (Figure~\ref{Harmonic_spectra}), MD‑based spectra (Figure~\ref{MLMD_spectra}), and PIMD, which accounts for quantum nuclear effects (Figure~\ref{MLMD_spectra} and ~\ref{Generability_MLMD}).

The main advantage of \MIR{} is its ensemble framework, which plays a dual role: improving robustness through model averaging and providing reliable uncertainty estimates that reflect sensitivity to chemical composition, bonding environments, and data coverage. For energies and dipole moments, uncertainty is naturally quantified at the molecular level through ensemble variance, while for forces we resolve uncertainty at the per-atom level. This enables element-resolved analysis of model confidence, revealing systematic trends in uncertainty across the periodic table and highlighting chemically challenging species, such as heavy main-group and transition-metal atoms. Although elements with higher representation in the training set generally exhibit lower uncertainty, this relationship is not strictly monotonic (refer to Figure~\ref{fig:uncertainty_force_qcml} to Figure~\ref{Dataset_diversity}). The reported values are averaged over many distinct coordination environments, and the same element may appear in chemically simple or highly complex bonding motifs, leading to substantial variation in local confidence. 
In this sense, uncertainty estimation goes beyond a statistical diagnostic and becomes a tool for probing the underlying chemistry encoded by the model.

In terms of harmonic IR spectra predictions, the results shown in Table~\ref{table_harmonic_ir} demonstrate that \MIR{} delivers accurate and reliable vibrational spectra across chemically diverse systems, despite being trained only on general-purpose energies, forces, and dipoles rather than spectroscopic observables. Across all three external test sets, the \MIR{} ensemble reproduces DFT harmonic frequencies and intensities with low mean absolute errors, highlighting \MIR{}'s ability to capture the underlying potential energy surfaces and dipole derivatives relevant for vibrational spectroscopy.
The strongest performance is observed for the QM7-x test set of organic molecules, where \MIR{} reaches near-DFT accuracy, owing to the close match between this chemical space and the training data, which minimizes extrapolation. Errors in this regime approach typical numerical uncertainties of harmonic DFT calculations of $\sim$10~cm$^{-1}$ \cite{howard2015assessing, pracht_comprehensive_2020, Pracht_2024}, underscoring the suitability of the model for high-throughput IR screening in conventional organic chemistry.

More challenging behavior is observed for the QCML-small and tmQM subsets, which include heavier elements and transition-metal-containing systems. The higher errors in these datasets reflect the increased complexity of their potential energy surfaces and dipole responses. Despite the increase in MAE to approximately 25~cm$^{-1}$, the overall accuracy remains reasonable, indicating that \MIR{} retains predictive capability even for chemically more complex systems.
Across QCML-small and tmQM, low-frequency modes exhibit slightly larger frequency errors, while high-frequency modes show increased intensity errors. These patterns likely reflect the greater sensitivity of low-frequency collective modes to small curvature errors in the potential energy surface, and the strong dependence of high-frequency stretching intensities on accurate dipole derivatives, particularly in electronically complex systems.

Our investigation of ensemble‑based uncertainty and its relationship to real error in the prediction of harmonic IR spectra was conducted for the first time, to the best of our knowledge. 
The observed correlation suggests that the ensemble provides a meaningful measure of confidence, helping to identify predictions that may be less reliable.
Across  all datasets, predictive uncertainty correlates strongly with the actual error for both frequencies and intensities (Figures S10-S12), with the correlation becoming particularly pronounced for high-error cases. Confusion-matrix analysis (Figure~\ref{fig:qcml_confusion_matrix}) further confirms that ensemble uncertainty effectively discriminates between reliable and unreliable predictions, providing a practical internal measure of confidence.

In sum, although direct comparison of harmonic spectra predictions to experiment is deferred due to the need for scaling factors \cite{Pracht_2024}, the strong agreement with DFT and reliable uncertainty estimates provide a solid foundation for quick spectra comparisons, when high-precision is not needed.

While harmonic spectra provide a useful baseline, experimentally measured vibrational spectra are governed by anharmonic finite-temperature dynamics. In both DFT-MD and ML-MD simulations, we observed systematic blue shifts of high-frequency vibrational peaks relative to the experiment. This reflects limitations of a purely classical treatment of nuclear motion and, in part, the known frequency overestimation associated with the hybrid PBE0 functional \cite{MACE_off_2025}.
However, ML-PIMD corrects these shifts, aligning high-frequency peaks closer to experimental positions and improving spectral agreement. Our calculations indicate that the magnitude of this correction varies by system, particularly for molecules where NQEs are strong. A notable example is 4-hydroxy-3-methoxybenzaldehyde, shown in Figure~\ref{MLMD_spectra}(d).
This improvement underscores the importance of NQEs in vibrational spectroscopy, especially for high-frequency regimes. The \MIR{} framework enables quantum dynamical simulations at a fraction of DFT's computational cost while maintaining predictive fidelity. These results demonstrate \MIR{}'s capacity to reproduce classical anharmonic dynamics and provide a practical approach for routinely incorporating NQEs in IR spectra simulations.

We also assessed \MIR{}’s ability to generalize across molecules with varying elemental representation in the training data (see Figure~\ref{Generability_MLMD}). While the IR spectra of molecules containing frequently occurring elements (e.g., C, H, N, O, Cl, I, Al, Zn) were predicted with high fidelity and low uncertainty, performance degraded for molecules containing elements that were sparsely represented in the training data, such as Ag, Li, and Ca. In these cases, the predicted spectra exhibit higher uncertainty, which increases progressively as elemental abundance in the training set decreases.
Ferrocene (d) represents an important exception in this series. Despite the moderate representation of Fe in the training data, the ML-PIMD predictions for ferrocene show markedly elevated force uncertainties (Table S5) and poor agreement with experiment. Trajectory-level analysis (Table~S6) indicates that the force uncertainty exceeds 10 eV/\AA\ as early as ~2 ps into the simulation, escalating further to over 100 eV/\AA\ at later steps. This highlights that even elements with moderate training coverage can pose challenges for organometallic systems with complex bonding patterns, and that high ensemble-derived uncertainty effectively signals regions where model predictions are less reliable.
The data in Table~S5 further indicate that when the mean force uncertainty exceeds approximately 0.1~eV/\AA, discrepancies between \MIR{} and experiment become more pronounced (e.g., frequency misalignment), as exemplified by the case shown in Fig.~\ref{Generability_MLMD}(c) with a value of 0.34~eV/\AA. However, rigorous calibration of this uncertainty metric, particularly the identification of a reliable and actionable threshold, would require a more extensive, dedicated study beyond the scope of the present work.

Overall, these findings reinforce that ensemble-derived uncertainty is a reliable indicator of model confidence: low uncertainty corresponds to robust predictions for well-represented elements, while elevated uncertainty identifies challenging cases, whether due to sparse data coverage (e.g., Ag, Li, Ca) or intrinsic chemical complexity (e.g., Fe in Ferrocene). Such insights suggest avenues for improvement, including targeted fine-tuning or transfer learning for underrepresented or chemically intricate species.

After discussing \MIR{} capabilities in terms of prediction and IR spectra simulations, we proceed to compare it with other existing trained models. 
However, a direct, quantitative comparison across models remains challenging, as these approaches are trained on different datasets, target properties, and electronic-structure reference levels, and therefore reported error metrics are not always directly comparable.
This includes models trained on OMOL25 dataset \cite{levine2025open}, universal models for atoms (UMA) \cite{wood2025family} and variants of the MACE-off \cite{kovacs2025mace} family, which have been explored for vibrational and dipole-related predictions. 

One particularly relevant point of reference is the recently introduced SO3LR \cite{kabylda_molecular_2025} potential, which is trained on DFT-PBE0 reference data with MBD corrections and explicitly targets long-range electrostatic interactions. We therefore performed a limited comparison on common benchmark datasets where published SO3LR results are available (see Table~\ref{tab:so3lr_comparison}). The two test sets, QM7-x \cite{donchev2021quantum} and TorsionNet500 \cite{rai2022torsionnet}, are adopted here and follow the exact evaluation protocols introduced in the original SO3LR work \cite{kabylda_molecular_2025}. More details on the two test sets are provided in the Methods section. 

In evaluating the QM7-x test dataset, \MIR{} delivers competitive results for force and dipole errors, although SO3LR achieves lower values. This difference may arise from more specialized training data and its focus on long-range electrostatics. On the chemically complex TorsionNet500 test set, \MIR{} outperforms SO3LR in force RMSE while maintaining similar dipole accuracy (Table~\ref{tab:so3lr_comparison}). Although both models incorporate many-body dispersion effects, \MIR{} employs a nonlocal MBD-NL formulation, whereas SO3LR uses only the standard version of MBD. This fundamental design difference contributes to variations in force and dipole accuracy, reflecting distinctions in electronic-structure targets beyond the models' capacities. Moreover, the comparison relies on RMSE values, which are consistently higher than MAEs. This indicates a small number of outliers which could be identified through ensemble uncertainty estimation.

\begin{table}[h]
\centering
\caption{Comparison of \MIR{} and SO3LR on common benchmark datasets (MAE values in parentheses are reported only for \MIR{}).}
\begin{tabular}{lccccc}
\hline
Dataset & Size & \# Atoms & Model & RMSE Force (eV/\AA) & RMSE Dipole (e\AA) \\
\hline
\multirow{2}{*}{QM7-x} 
& \multirow{2}{*}{10k} 
& \multirow{2}{*}{6--23} 
& \MIR{} & 0.101 (0.038) & 0.094 (0.041) \\
& & & SO3LR & 0.069 & 0.031 \\
\hline
\multirow{2}{*}{TorsionNet500} 
& \multirow{2}{*}{12k} 
& \multirow{2}{*}{13--37} 
& \MIR{} & 0.066 (0.038) & 0.098 (0.060) \\
& & & SO3LR & 0.088 & 0.061 \\
\hline
\label{tab:so3lr_comparison}
\end{tabular}
\end{table}

Overall, this comparison highlights that while specialized models trained at fixed high-level reference theories may achieve lower or comparable errors on narrowly defined benchmarks, \MIR{} offers a complementary and more general-purpose alternative. By combining competitive accuracy, various dispersion treatments, uncertainty quantification from ensembles, and strong transferability across chemically diverse systems, \MIR{} provides a flexible foundation for vibrational spectroscopy and molecular dynamics beyond the specific regimes targeted by existing potentials.

Finally, our approach offers a significant advancement in both accuracy and scalability, against traditional methods such as DFT. DFT-based AIMD and PIMD simulations provide high-fidelity IR spectra but are computationally expensive, especially for large systems or long trajectories. In contrast, with our \MIR{} model, spectra can be predicted orders of magnitude faster, reducing computational time from thousands of CPU hours to just minutes or hours on a single GPU, while maintaining comparable accuracy. This efficiency makes it feasible to explore larger systems, longer simulation times, and a broader range of molecular configurations.

\section{Conclusion}\label{conclusion}
We introduced \MIR{}, an uncertainty-aware foundation model for molecular infrared spectroscopy. The framework combines the MACE architecture, ensemble learning, and large-scale hybrid-DFT reference data with different dispersion treatments to deliver accurate predictions of energies, forces, dipole moments, and IR spectra across chemically diverse systems. On established benchmarks, the model achieves competitive accuracy while maintaining broad transferability beyond its training domain. Crucially, the ensemble formulation provides chemically interpretable uncertainty estimates that track prediction reliability at both molecular and per-atom resolution, enabling systematic identification of challenging chemical environments without external reference calculations.

Accurate harmonic spectra provide a reliable baseline for vibrational analysis, while coupling \MIR{} with molecular dynamics extends the framework to fully anharmonic, finite-temperature simulations. ML-MD and ML-PIMD trajectories reproduce experimental spectral trends, with the latter demonstrating that inclusion of NQEs improves agreement in the high-frequency regime. These results establish a consistent framework that spans static harmonic analysis and quantum dynamical spectroscopy at a fraction of the cost of DFT.

By combining predictive accuracy with uncertainty awareness and quantum-dynamical capability, \MIR{} establishes a practical framework for reliability-aware spectral simulations at scale. This foundation naturally enables future directions such as inverse spectral design, automated reaction monitoring, and integration into high-throughput screening pipelines, where fast and trustworthy vibrational predictions are essential.

\section{Methods}\label{method}
\subsection{Machine learning models}

In this work, we employed the MACE architecture~\cite{Batatia2022Design, MACE} to predict total energies, atomic forces, and molecular dipole moments from \textit{ab initio} reference data. All models used in \MIR{}, including those forming the ensemble, share the same underlying architectural design described below, differing only in training data splits and, where applicable, the inclusion of dispersion corrections. This ensures that uncertainty estimates arise from data- and optimization-induced variability rather than from heterogeneous model classes.  

For predicting total energies and atomic forces (MACE-EF), we trained MACE models of varying sizes, characterized by (i) the number of channels (\texttt{num\_channels}, defining the dimensionality of node features), (ii) the angular-momentum cutoff (\texttt{Max\_L}). MACE represents messages with features labeled by the angular-momentum quantum number $L$: scalar properties (e.g., total energy) transform as $L=0$, whereas vectorial properties (e.g., forces, dipole moments) require higher-order ($L>0$) equivariant features. The distinction between \textit{"Small"}, \textit{"Medium"}, and \textit{"Large"} models is summarized in Table~\ref{table_mace_architecture}. All energy–force models (MACE-EF) use a cutoff radius of 5.0~\AA, two interaction layers, and a body order of 3, and were trained using \texttt{mace} version~0.3.10.

\begin{table}[h]
\caption{MACE model's parameters trained for energy and force prediction.}
\label{table_mace_architecture}
\begin{tabular}{@{}lll@{}}
\toprule
\textbf{Model size} & \textbf{num\_channels} & \textbf{Max\_L} \\
\midrule
Small               & 128                  & 1 \\
Medium              & 196                  & 1 \\
Large               & 256                  & 2 \\
\botrule
\end{tabular}
\end{table}

In addition to the energy and force models, we trained a separate MACE model to predict molecular dipole moments (MACE-D), which are essential for IR spectral analysis.  The primary architectural difference lies in the readout layer: instead of summing scalar atomic energies, it outputs a vectorial quantity per atom, which is then summed to obtain the total dipole moment of the molecule. Multiple variants were systematically evaluated, differing in channel dimensionality, angular-momentum cutoff, and numerical precision, as summarized in Table~\ref{table_dipole_training_curve}. For the \MIR{} ensemble, we selected a large-capacity MACE-D model with 128 channels, and also include additional models with 64 and 92 channels to enable robust and more efficient ensemble predictions. All models employ $L=1$ equivariant messages with a 5.0~\AA\ cutoff, ensuring accurate and rotationally consistent dipole predictions.

All training and inference procedures were conducted on AMD MI250x GPUs, ensuring high throughput for large-scale model development and deployment.

\subsection{DFT computational details}
All DFT calculations are in agreement with the settings used to create QCML database~\cite{ganscha_qcml_2025}. More precisely, the calculations were performed using the all-electron numeric-atom-centered orbital code FHI-aims \cite{blum2009ab, havu2009efficient, Xinguo/implem_full_author_list, levchenko2015hybrid}. The hybrid Perdew-Burke-Ernzerhof exchange-correlation functional (PBE0) \cite{PBE, PBE0} was used for the calculations. Further computational settings included the standard FHI-aims tier-2 basis sets and "tight" grid settings, the zeroth-order regular approximation to account for scalar relativistic effects \cite{lenthe1993relativistic}.
Structure optimization was carried out using the Broyden–Fletcher–Goldfarb–Shanno (BFGS) minimizer \cite{nocedal_numerical_2006}, with a convergence limit of 0.01~eV/Å for the maximum atomic force amplitude.
AIMD simulations were performed using the conventional FHI-aims software, while the geometry optimization and harmonic calculations were conducted using the Atomic Simulation Environment (ASE) \cite{ase-paper} FHI-aims calculator.

\subsection{QCML dataset}
In this work, we utilized the QCML dataset \cite{ganscha_qcml_2025}, which represents an initial step toward building a universal quantum chemistry database. The dataset is constructed from 17.2 million unique chemical graphs derived from both known molecular fragments and synthetically generated structures. From these graphs, a total of 14,678 million molecular conformations are sampled across a wide temperature range (0–1000 K), and their properties are initially computed using semi-empirical methods.
A randomly selected subset of 33.5 million conformers is then subjected to more accurate quantum mechanical calculations using DFT. This DFT-level dataset includes a wide range of quantum chemical properties, such as total energies, atomic forces, multipole moments, and matrix-based quantities like the Hamiltonian.

The 33.5 million DFT-evaluated conformers span neutral and charged species across various spin multiplicities. For our purposes, we retained only the neutral molecules with a spin multiplicity of 1. We further removed outlier data points from the data set using the flag mentioned in the QCML work.
This filtering process resulted in approximately 15.9 million clean, high-quality DFT structures. From this pool, 100,000 structures were randomly sampled to form the validation and test sets (QCML-large). The remaining structures were divided into three subsets of approximately 10 million structures each. 
Subsequently, the \MIR{} ensemble was constructed using the independent random splits  for each member.
Additionally, ensemble members were trained with different van der Waals treatments, non-local version of MBD, DFT-D4, or no explicit dispersion, allowing the capture of distinct interaction regimes while retaining a consistent neural network backbone.

\subsection{Test set construction}
We constructed five different test sets to evaluate the \MIR{} models in diverse chemical spaces.

\textbf{QCML-large test set:} 100,000 structures randomly sampled from neutral, singlet structures in reference QCML database calculated by DFT.

\textbf{QCML-small test set:} This subset was sampled from the QCML-large test data, explicitly targeting molecules containing non-organic (“rare”) elements. H, C, N, and O were excluded from consideration, and all remaining elements were treated as rare. Each structure was parsed to determine the unique set of elements present, and structures were classified according to the number of rare elements: exactly one, exactly two, or more than two. Structures containing exactly one rare element were treated as primary candidates, while structures with exactly two rare elements were used as fallback candidates. For each rare element, up to ten structures were selected, prioritizing primary candidates and supplementing with fallback structures when necessary. Selection was performed independently for each rare element, and the final test set was formed as the union of all selected structures. In total, the QCML-small test set spans 76 elements and 672 structures, encompassing a wide diversity of chemically relevant species. 

\textbf{QM7-x subset:} To benchmark performance in conventional organic chemical space, we used the QM7-x dataset (H, C, N, O, S, Cl) \cite{donchev2021quantum} following the test split introduced in the SO3LR work \cite{kabylda_molecular_2025}. The original SO3LR test set contains 10,000 structures computed at the PBE0 level with MBD dispersion corrections. From this published test split, we randomly selected 300 structures restricted to H, C, N, and O compositions to construct a representative subset suitable for our study. For consistency with our workflow, this subset was recomputed at the hybrid PBE0/tight level of theory using the FHI-aims software. This procedure preserves the statistical characteristics of the SO3LR benchmark while enabling direct comparison under our computational settings.

\textbf{TorsionNet500:} We adopted the TorsionNet500 test set introduced in \cite{rai2022torsionnet} and used in the SO3LR study \cite{kabylda_molecular_2025}. This dataset contains small molecules composed of H, C, N, O, F, S, and Cl and is designed to probe torsional energy profiles in chemically challenging conformational spaces. We used the original SO3LR test set without modification, retaining its reference properties computed at the PBE0 level with MBD dispersion corrections to ensure direct comparability with reported SO3LR results.

\textbf{tmQM subset:} To assess performance on transition-metal-containing systems, we created a test set consisting of 50 structures with transition-metal atoms extracted from reference \cite{balcells_tmqm_2020}. All structures additionally include H, C, N, and O atoms as required. This subset evaluates the model’s ability to generalize to coordination chemistry and transition-metal complexes. This subset was recomputed at the hybrid PBE0/tight level of theory using the FHI-aims software.

Collectively, these test sets (QCML-large, QCML-small, QM7-x subset, TorsionNET500 and tmQM subset) provide a diverse and chemically representative evaluation benchmark spanning rare-element chemistry, standard organic molecules, and transition-metal complexes.

\subsection{Harmonic IR spectra calculation} 
Harmonic IR spectra were computed using the harmonic approximation, where the potential energy surface is approximated as a quadratic function near the optimized geometry. Vibrational frequencies are obtained by diagonalizing the mass-weighted Hessian matrix of second derivatives with respect to atomic coordinates \cite{harmonic_1, Harmoni_2, AIMD_paper}. IR intensities were computed from the derivatives of the molecular dipole moment with respect to mass-weighted normal mode coordinates. The integrated IR absorption intensity \( A_\nu \) for a normal mode \( p \) is given by \cite{AIMD_paper}:

\begin{equation}
A_\nu = \frac{1}{4\pi \varepsilon_0 }\frac{N_A \pi}{3 c^2 } \left( \frac{\partial \boldsymbol{\mu}}{\partial Q_p} \right)^2,
\end{equation}
where \( Q_p \) is the mass-weighted normal coordinate of mode \( p \), \( \boldsymbol{\mu} \) is the molecular dipole moment vector, \( N_A \) is Avogadro’s number, \( c \) is the speed of light, and \( \varepsilon_0 \) is the vacuum permittivity. All calculations of harmonic IR spectra were performed using the ASE calculator and the structures were displaced with 0.002 \AA.

\textbf{Harmonic benchmark datasets construction:} The following protocol was applied to generate and filter the benchmark datasets. For each test set, the initial molecular structures were first pre-relaxed using the first \texttt{MACE-EF} model from the \MIR{} ensemble to ensure physically reasonable geometries and numerical stability. Geometry optimization and reference harmonic IR spectra were then computed using DFT. Structures that failed to converge during this step were excluded from the benchmark.

After applying these criteria and additional filtering, removing molecules with mean absolute errors in vibrational frequencies exceeding 200~cm$^{-1}$, dimensionality mismatches, or missing vibrational modes, the final benchmark set consisted of 472 molecules from QCML-small, 297 molecules from QM7-x, and 34 molecules from tmQM, yielding a total of 803 molecules for harmonic IR analysis. For more details on the dataset statistics, see Section~1.0 in the Supplementary Information.

\subsection{MD simulations} 
The DFT-based AIMD (DFT-MD) simulations for IR spectra generation were conducted in two stages. First, the system was equilibrated for 4 ps at the target temperature using the Berendsen thermostat with a relaxation time of 0.1 ps~\cite{berendsen_molecular_1984}. This was followed by a production run of 50 ps using the Nosé–Hoover thermostat, employing a thermostat mass equivalent to 4000 cm$^{-1}$~\cite{nose_unified_1984, hoover_canonical_1985}. Only the trajectory from the Nosé–Hoover stage was used in the final IR spectra calculations. For the DFT‑MD simulations of molecule (i) in Figure~\ref{MLMD_spectra}, we performed a 20~ps production run due to the high computational cost. 

ML-based spectra were obtained by averaging results from three independent trajectories, each generated by a different model in the ensemble. Prediction uncertainty was quantified as the standard deviation across these trajectories.

All ML-MD simulations were performed using a Langevin thermostat~\cite{ceriotti_langevin_2009} with a friction coefficient of 0.01, as implemented in ASE version 3.22.1~\cite{ase-paper}. First, 5 ps of each trajectory was used for thermalization, while the following 50 ps were used for the IR simulation.

All ML-PIMD simulations were calculated using the TRPMD-GLE \cite{rossi2018fine} formalism with 32 beads, interfaced through i-PI \cite{litman2024ipi} version 3.1.7 to communicate with the force evaluation engines. The GLE parameters were generated using the GLE4MD website \cite{rossi2018fine}. Similarly, to ML-MD simulations, initial 5 ps served for thermalization, and following 50 ps trajectory was utilized for the IR simulation.

Several PIMD approaches were evaluated, including centroid molecular dynamics (CMD) \cite{cao1994formulation, jang1999derivation}, ring‑polymer molecular dynamics (RPMD) \cite{craig2004quantum, braams2006short}, and thermostatted variants such as TRPMD \cite{rossi2014remove}. The TRPMD‑GLE method was selected because it showed the best agreement with experimental IR spectra of methanol and is the most established of the tested approaches.

All MD simulations employed a time step of 0.5 fs, with the temperature maintained at 300 K unless stated otherwise.

\subsection{MD simulations-based IR spectra calculation}
IR spectra were obtained from MD trajectories by evaluating the autocorrelation function of the time derivative of the dipole moment, $\dot{\mu}$. This is expressed as \cite{AIMD_paper}:

\begin{equation} I_{IR}(\omega) \propto \int_{-\infty }^{+\infty}\left\langle \dot{\mu}\left( \tau \right) \dot{\mu}\left( \tau + t \right) \right\rangle_{\tau}e^{-i\omega t}dt .\label{eq:IR-intensity} \end{equation}

In this work, all auto-correlation functions were computed using the Wiener–Khinchin theorem \cite{wiener_generalized_1930}. To improve the quality of the resulting spectra, a Hann window function~\cite{blackman_measurement_1958} and zero-padding were applied prior to the Fourier transform. A maximum correlation time of 1000 fs was used throughout. All spectral processing was carried out using a slightly modified version of the auto-correlation code from SchNetPack~\cite{schutt_schnetpack_2019}.

\backmatter

\bmhead{Acknowledgements}
N.B., S.B., P.R., and M.A.L.M. acknowledge the funding from Horizon Europe MSCA Doctoral network grant n.101073486, EUSpecLab, funded by the European Union.
O.K. and P.R. have received funding from the European Union – NextGenerationEU instrument and are funded by the Research Council of Finland (grant numbers 348179, 346377, and 364227).
We acknowledge CSC, Finland for awarding access to the LUMI supercomputer, owned by the EuroHPC Joint Undertaking, hosted by CSC (Finland) and the LUMI consortium through CSC, Finland, extreme-scale project ALVS.
The authors also gratefully acknowledge additional computational resources provided by CSC – IT Center for Science, Finland, and the Aalto Science-IT project.

\section*{Declarations}
The authors report no conflicts of interest. The trained models, scripts, and a tutorial on how to use the models are available in the HuggingFace repository (\url{https://huggingface.co/nitbha007/MACE4IR}), while the dataset used for training, the AIMD and PIMD simulation data, and all input files for the harmonic and MD-based spectra calculations are available on Zenodo (doi: 10.5281/zenodo.16761021 and doi: 10.5281/zenodo.16920067).
N.B. created the workflow and proceeded the calculation.
O.K., S.B., P.R., and M.M. supervised the work.
All authors contributed to the manuscript. 

\bmhead{Supplementary information}
The Supplementary  information contains:~\\
Figure S1 and S2 – Elemental distribution in the filtered 10M QCML dataset used to train the second and third ensemble model respectively. Figure S3 and S4 – Element-resolved heatmaps of ensemble force uncertainty for the QM7-x and tmQM test subsets. Figure S5 and S6 – Element-resolved heatmaps of ensemble energy and dipole uncertainty for the QCML test subset. Figure S7, S8, and S9 – Distributions of ensemble uncertainty for QCML energy, forces, and dipole moments. Figure S10 – Correlation between ensemble uncertainty and spectral prediction error for the QCML harmonic benchmark. Figure S11 and S12 – Uncertainty–error correlation analysis for harmonic spectra on the QM7-x and tmQM test subset respectively. Figure S13 and S14 – Analysis of confusion matrices for harmonic spectra in the QM7-x and tmQM test subset, respectively. Figure S15 – Representative molecules for which harmonic spectra calculations failed due to DFT convergence issues or large DFT–ML discrepancies. \\

Table S1 – Dipole moment scaling study showing accuracy–cost tradeoffs across model architectures. Table S2 – Dipole moment prediction accuracy across multiple benchmarks for different model architectures. Table S3 – Spectral similarity metrics between experimental, DFT, ML-AIMD, and ML-PIMD IR spectra. Table S4 – Experimental vs ML-PIMD spectral agreement across representative molecules.
Table~S5 - Mean and standard deviation of predicted energy, force, and dipole uncertainties ($E_{\sigma}$, $F_{\sigma}$, $\mu_{\sigma}$) along the first PIMD trajectory for each molecule.  
Table~S6 - Trajectory-level force uncertainty ($F_{\sigma}$) diagnostics for Ferrocene, showing threshold crossings and total simulation time. \\

A detailed description of the harmonic IR benchmarking protocol is provided in Section~1.0, while the IR preprocessing and spectral similarity analysis are described in Section~2.0 in Supplementary Information.

\bibliography{sn-bibliography}% common bib file

\end{document}